\documentclass[fleqn,usenatbib]{mnras}

\usepackage{newtxtext,newtxmath}
\usepackage[T1]{fontenc}

\DeclareRobustCommand{\VAN}[3]{#2}
\let\VANthebibliography\thebibliography
\def\thebibliography{\DeclareRobustCommand{\VAN}[3]{##3}\VANthebibliography}

\usepackage{graphicx}	
\usepackage{amsmath}	
\usepackage{adjustbox}

\title[Updated and forecast constraints on $\Omega_k$, $M$ and $r_d$]{Revisiting model-independent constraints on spatial curvature and cosmic ladders calibration: updated and forecast analyses}

\author[A. Favale, A. G\'omez-Valent \& M. Migliaccio]
{Arianna Favale$^{1,2,3,4,5}$\thanks{afavale@roma2.infn.it}, Adri\`a G\'omez-Valent$^{4,5}$\thanks{agomezvalent@icc.ub.edu} and Marina Migliaccio$^{1,3}$\thanks{migliaccio@roma2.infn.it}
 \\
$^{1}$Dipartimento di Fisica, Università di Roma Tor Vergata, via della Ricerca Scientifica, 1, 00133, Roma, Italy\\
$^{2}$Dipartimento di Fisica, Università di Roma La Sapienza, Ple Aldo Moro 2, 00185, Roma, Italy\\
$^{3}$INFN, Sezione di Roma 2, Università di Roma Tor Vergata, via della Ricerca Scientifica, 1, 00133 Roma, Italy\\
$^{4}$Departament de Física Quàntica i Astrofísica (FQA), Universitat de Barcelona (UB), c. Martí i Franqués, 1, 08028 Barcelona, Catalonia, Spain\\ 
$^{5}$Institut de Ciències del Cosmos (ICCUB), Universitat de Barcelona (UB), c. Martí i Franqués, 1, 08028 Barcelona, Catalonia, Spain
}

\begin{document}
\label{firstpage}
\pagerange{\pageref{firstpage}--\pageref{lastpage}}
\maketitle

\begin{abstract}

In recent years, model-independent approaches have gained increasing attention as powerful tools to investigate persistent tensions between cosmological observations and the predictions of the standard $\Lambda$CDM model. Notably, recent data from the DESY5 Type Ia Supernovae (SNIa) sample and the latest Baryon Acoustic Oscillation (BAO) measurements from the DESI collaboration challenge the validity of the cosmological constant, and under the assumption of standard pre-recombination physics, they still remain in tension with the SH0ES local distance ladder measurements.
Building on our previous work, Mon. Not. Roy. Astron. Soc. 523 (2023) 3, 3406-3422, we present a follow-up analysis of the model-independent calibration of both the local and inverse distance ladders using cosmic chronometers (CCH) data and the Gaussian Processes technique. We constrain the SNIa absolute magnitude, $M$, and the comoving sound horizon at the baryon-drag epoch, $r_d$, while simultaneously deriving a measurement of the spatial curvature parameter, $\Omega_k$, using CCH with DESY5 and DESI DR1 and DR2 data releases. Our results show that this data combination is compatible with a flat universe at $\sim1.7\sigma$, with $\Omega_k = -0.143 \pm 0.085$, indicating a weaker compatibility than that observed with SNIa from Pantheon+, while the ladders calibrators read $M=-19.324_{-0.095}^{+0.092}$ and $r_d = (144.00^{+5.38}_{-4.88}$) Mpc. Although current uncertainties limit the precision of our constraints and prevent us from arbitrating the Hubble tension, it is nevertheless instructive to explore the constraining power of our methodology with future SNIa, CCH, and BAO observations from surveys such as Vera C. Rubin Observatory - LSST, Euclid, and DESI. Thus, for the first time, we present a forecast analysis for the triad $(M, \Omega_k, r_d)$. Our results indicate that, in an optimistic scenario, upcoming data will improve agnostic constraints on the ladder calibrators - $M$ by $\sim$54\%, $r_d$ by $\sim$66\% - which enable us to constrain the Hubble parameter, $H_0$, at a 2\% level. Precision on $\Omega_k$ will increase by $\sim50\%$. Our analysis outlines which improvements in future data -- whether in quality, quantity, or redshift coverage -- are likely to have the greatest impact on tightening these constraints.

\end{abstract}

\begin{keywords}
cosmological parameters -- dark energy -- distance scale -- cosmology: observations.
\end{keywords}



\section{Introduction}

With the advent of increasingly precise observations, the standard model of cosmology, namely the $\Lambda$CDM, is facing growing scrutiny and active debate, which strongly challenge some of its pillars and fundamental assumptions. The Hubble tension -- a $\gtrsim5\sigma$ mismatch between local measurements of the Hubble-Lema\^itre constant, $H_0$, and the value inferred by early universe observations under the $\Lambda$CDM model -- remains one of the most prominent challenges in modern cosmology. This, together with a broader pattern of inconsistencies, may call for a reassessment of our current concordance model (see \cite{Perivolaropoulos:2021jda} and \cite{DiValentino:2025sru} for dedicated reviews).  $H_0$ is one of the six parameters of the $\Lambda$CDM model. Having an accurate measurement of this quantity is crucial for an unbiased estimation of the total energy content of the universe via the Friedmann equation, its age, and also for a correct determination of cosmic distances. 
While the Cosmic Microwave Background (CMB) data from {\it Planck} \citep{Aghanim:2018eyx}, the Atacama Cosmology Telescope (ACT) \citep{AtacamaCosmologyTelescope:2025blo} and the South Pole Telescope (SPT) \citep{SPT-3G:2025bzu} favor a value of $H_0 \sim 67$ km/s/Mpc, many local distance ladder measurements prefer values closer to $H_0 \sim 73$ km/s/Mpc \citep{Riess:2021jrx, Riess:2024vfa}. These local estimates crucially rely on the standardization of standard candles, such as Supernovae of Type Ia (SNIa). The intrinsic luminosity -- or absolute magnitude, $M$ --  of nearby SNIa ($z \lesssim 0.01$) can be calibrated in galaxies that host Cepheid variables, which themselves are tied to absolute scales using, for example, Gaia parallaxes in the Milky Way. By extending the ladder up to the Hubble flow, i.e., up to $z\sim 0.15$, $H_0$ can be measured in a highly model-independent way, assuming only the validity of the cosmological principle. However, the direct distance ladder is subject to several sources of systematic errors. The current discrepancy in the determination of $H_0$ has stimulated several studies aimed at reassessing the robustness of this method. For instance, some works have highlighted potential systematics in the intermediate rung of the local distance ladder, which may bias the current expansion rate to higher values, see, e.g., \cite{Mortsell:2021tcx, Desmond:2019ygn, Desmond:2020wep, Hogas:2023pjz, Wojtak:2024mgg}. Consolidations on the use of anchors are also ongoing \citep{Riess:2024vfa, Freedman:2024eph}. Other standard candles, like, e.g., the Tip of the Red Giant Branch \citep{Freedman:2023jcz, Freedman:2024eph}, Mira variables \citep{Huang:2023frr} or J-regions of the asymptotic giant branch \citep{Li:2024yoe,Lee:2024qzr} are viable alternatives for estimating $H_0$ without relying on early-time processes and possibly bypassing some of the Cepheid-related systematics. Notably, some of these alternative studies lead to smaller values of $H_0$, see, e.g., \cite{Freedman:2024eph,Lee:2024qzr}.

Conversely, one can approach the problem using objects of known size in our universe that serve as standard rulers. This is the case of $r_d$, the comoving sound horizon at the baryon drag epoch, which characterizes the typical size of the Baryon Acoustic Oscillations (BAO) when baryons decoupled from photons\footnote{The horizon size at the time of matter-radiation equality might also become a viable alternative, once we have more data points, see \cite{DESI:2025euz}.}. This scale is imprinted in both the CMB temperature anisotropies and in the large-scale distribution of galaxies. A prior on $r_d$ derived from CMB or a prior on the physical baryon matter density from Big Bang Nucleosynthesis can be used to calibrate anisotropic BAO data, allowing for a measurement of $H_0$  under the assumption of standard pre-recombination physics. These calibrated BAO absolute distances can be used to anchor the relative distances from SNIa independently of the local distance ladder and enable the inference of $H_0$ (and $M$) by extending the distance-redshift relation down to $z=0$. This methodology, known as the inverse distance ladder, leads to $H_0$ values closer to the CMB result when anisotropic (3D) BAO data are employed in the analyses \citep{Aubourg:2014yra,Cuesta:2014asa,Addison:2017fdm,DES:2017txv,Feeney:2018mkj,DES:2024ywx}.
Standard pre-recombination physics sets $r_d$ at approximately 147 Mpc. When this value is used to calibrate anisotropic BAO data, which in turn are combined with CMB and uncalibrated SNIa, one gets values of the absolute magnitude of $M\sim-19.40$ (see, e.g. \cite{Gomez-Valent:2022hkb}). This result is still in $\sim5\sigma$ tension with that obtained from the local distance ladder method, $M^{R22} = -19.253\pm0.027$ \citep{Riess:2021jrx}. This discrepancy becomes clear by simply studying the compatibility of uncalibrated SNIa and 3D BAO data, which sets the calibrators of the direct and inverse distance ladders to lie in a quite narrow degeneracy band in the $M-r_d$ plane (see \cite{Gomez-Valent:2023uof}).
The mismatch between the overlapping region of the SNIa+BAO degeneracy band and the values of $M^{R22}$ and the {\it Planck}-inferred $r_d$, frames the Hubble tension as a tension in the calibrators of the local and inverse distance ladder \citep{Gomez-Valent:2023uof, Perivolaropoulos:2024yxv}. It is therefore essential to measure these two quantities with different and independent techniques, which are also independent of the underlying cosmological model. This way, estimates of $M$ and $r_d$ can be employed to identify the presence of possible systematic errors in the ladders, and also to evaluate and discriminate among alternatives to the $\Lambda$CDM model in view of the Hubble tension.

For instance, dynamical dark energy models are now at the center of intensive studies, since current data from CMB, BAO and SNIa prefer a highly non-trivial evolution of the dark energy density at $\sim 3-4\sigma$ C.L., depending on the concrete SNIa sample employed \citep{DESI:2025zgx,DESI:2025fii}, with a probability of phantom divide crossing of 96.21\%-99.97\% \citep{Gonzalez-Fuentes:2025lei}, see also \cite{Ozulker:2025ehg}\footnote{For some previous hints of dynamical dark energy see, e.g., \cite{Alam:2004jy,Alam:2003fg,Sahni:2014ooa,Salvatelli:2014zta,Sola:2015wwa,Sola:2016jky,SolaPeracaula:2016qlq,SolaPeracaula:2017esw,Zhao:2017cud,Sola:2017znb,SolaPeracaula:2018wwm}.}. These late-time dynamical dark energy models lead to values of $H_0$ at $>5\sigma$ tension with the SH0ES measurement. However, including SH0ES as a prior with CMB+SNIa+BAO data and considering also the presence of early dark energy reduces the preference for dynamical dark energy at low redshift down to 1.5-2.4$\sigma$ level \citep{Poulin:2024ken,Pang:2025lvh,Adi:2025hyj}. Indeed, it is well-known that no late-time dynamical dark energy can solve the Hubble tension in the presence of anisotropic BAO data \citep{Sola:2017znb,Krishnan:2021dyb,Lee:2022cyh,Keeley:2022ojz,Gomez-Valent:2023uof}. Hence, the assessment of the robustness of SH0ES measurement is crucial also for a solid quantification of the evidence for dynamical dark energy. On the other hand, angular (2D) BAO allows for late-time solutions to the Hubble tension \citep{Bernui:2023byc,Akarsu:2023mfb, Gomez-Valent:2023uof,Gomez-Valent:2024tdb,Dwivedi:2024okk,Gomez-Valent:2024ejh}, which calls into question the reliability of conclusions drawn from different BAO measurements. Indeed, 2D and 3D BAO data are found to be in tension at a non-negligible level \citep{Camarena:2019rmj,Favale:2024sdq}.

While the full picture remains uncertain, this evolving landscape has motivated a variety of model-independent analyses that aim to shed light on these conundrums, spanning approaches from cosmography to non-parametric techniques such as Gaussian Processes (GP) and neural networks (see, e.g., \cite{Yu:2017iju,Gomez-Valent:2018hwc,Haridasu:2018gqm,  Collett:2019hrr, Benisty:2022psx, Cao:2023eja, Renzi:2020fnx, Mukherjee:2024ryz, Mukherjee:2025ytj, Ling:2025lmw}).

In this work, we give the CCH data the analogue role that Cepheids and BAO play in building the local and inverse distance ladders, respectively; that is, we use cosmic clocks as independent calibrators of $M$ and $r_d$. We do so by setting a model-independent framework which, notably, is also independent of the main drivers of the Hubble tension -- the first rungs of the cosmic distance ladder employed by SH0ES and the CMB data. At the same time, our approach enables constraints on the spatial curvature, which enters the geometrical pre-factor in the definition of the luminosity distance. Within the $\Lambda$CDM framework, CMB data from the {\it Planck} 2018 TT,TE,EE+lensing likelihood prefer a closed universe at $< 2\sigma$ C.L., $\Omega_k=-0.0106\pm 0.0065$ \citep{Aghanim:2018eyx,Handley:2019tkm,DiValentino:2019qzk}. However, the compatibility with a flat universe is recovered as soon as data on BAO, SNIa, the full-shape galaxy power spectrum or CCH are added on top of CMB \citep{Aghanim:2018eyx,Efstathiou:2020wem,Vagnozzi:2020rcz,Vagnozzi:2020dfn}, or when using only CMB data from ACT \citep{ACT:2020gnv}. Moreover, the final (PR4) {\it Planck} data release leads to $\Omega_k=-0.012\pm 0.010$ \citep{Tristram:2023haj}, again compatible with the flat scenario, see also \cite{Asorey:2025hgx}. Within the CCH calibration technique, here, we constrain $\Omega_k$ using a quite model-independent approach, which relies on the validity of the cosmological principle and a metric description of gravity. The original idea was proposed by \cite{Sutherland:2012ys} and applied for the first time by \cite{Heavens:2014rja}. 
We already adopted this methodology in \cite{Favale:2023lnp}, and several works in the literature developed the method through alternative data sets and statistical tools, see, e.g., \cite{Verde:2016ccp, Haridasu:2018gqm, Yang:2020bpv, Dhawan:2021mel, Gomez-Valent:2021hda, Zhang:2023eup, Qi:2023oxv, Jiang:2024xnu}.

Our approach leverages Gaussian Processes, a technique now widely used in cosmology due to its property of relying on very minimal assumptions that enable agnostic estimation of functions of interest. We employ it to reconstruct the Hubble function, thereby fixing the background history from which SNIa and BAO cosmic distances can be inferred. A joint analysis using these three data sets enables us not only to put constraints on the triad $(M, \Omega_k, r_d)$ but also to obtain an estimate of the Hubble parameter, $H_0$, through a cosmographic approach.

In particular, in this paper, we present a follow-up analysis of our previous work in \cite{Favale:2023lnp}. We do so by substituting Pantheon+ SNIa \citep{Scolnic:2021amr} with the more recent DESY5 sample \citep{DES:2024jxu} and using the anisotropic BAO data from the data releases DR1 and DR2 of DESI \citep{DESI:2024uvr,DESI:2025zgx}. Assessing the impact of these new data is crucial in light of the current claims and ongoing debates. Additionally, motivated by our previous work in \cite{Favale:2024sdq}, which highlighted a $2.5-4.6\sigma$\footnote{The range of this tension depends on the concrete SNIa data sets employed in the analysis (Pantheon+ or DESY5). In particular, we found the largest mismatch using DESY5 data. See \cite{Favale:2024sdq} for further details.} discrepancy between 2D BAO and 3D BAO from DESI DR1, we explore how the replacement of anisotropic measurements with angular measurements affects the results.
Moreover, we also consider it important to take a step forward and look at the constraining power that future data will have in the context of this model-independent methodology.
Of course, minimal assumptions translate into larger uncertainties, which at present represents its main limitation.
However, current and upcoming surveys -- such as the Euclid ESA mission \citep{EUCLID:2011zbd}, the Vera Rubin Observatory’s Legacy Survey of Space and Time (LSST) \citep{LSSTScience:2009jmu}, the Roman Space Telescope \citep{2015arXiv150303757S}, the Simons Observatory \citep{SimonsObservatory:2018koc}, and the LiteBIRD mission \citep{2023PTEP.Litebird}  -- are expected to deliver transformative improvements in both statistical precision and control of systematics. It is therefore timely to assess whether model-independent methodologies can achieve competitive performance relative to model-dependent constraints in light of these forthcoming data.

This paper is organised as follows. In Sec. \ref{sec:data}, we present the current low-$z$ data on CCH, SNIa and BAO employed in the first part of the analysis, along with the description of the mock data sets built to perform the forecast analysis. The methodology is detailed in Sec. \ref{sec:method}, with a description of the GP technique and the joint analysis framework developed to calibrate the ladders and measure $\Omega_k$. This is done for different data set combinations. Results of these analyses are presented in Sec. \ref{sec:res} both for current data and forecasts. In Sec. \ref{sec:conclusions}, we finally draw our conclusions.

\section{Data sets}\label{sec:data}

\subsection{Data from current surveys}\label{sec:data_curr}
In this section, we present the state-of-the-art data on CCH, SNIa and BAO. These data sets have been updated from our previous work  \citep{Favale:2023lnp}.

\subsubsection{CCH}\label{sec:cch_curr}
Cosmic chronometers stand out as one of the ideal probes for model-independent analyses, as they only rely on minimal assumptions in the local environment of the stars, namely the validity of General Relativity and standard stellar physics.
As first proposed by \cite{Jimenez:2001gg}, the CCH method is based on the fact that, in a Friedmann-Lema\^itre-Robertson-Walker (FLRW) universe, the Hubble parameter can be expressed in terms of the differential ageing of the universe, $dt$, and as a function of $z$ via the relation

\begin{equation}
     H(z) = \frac{-1}{1+z}\frac{dz}{dt}\,.
 \end{equation}
The ratio $dz/dt$ is not directly observable, but it can be inferred by identifying astrophysical objects whose age evolution across redshifts is well understood \footnote{In the very long term, additional model-independent measurements of $H(z)$ inferred from redshift-drift observations might be possible thanks to future facilities such as the Extremely Large Telescope and the Square Kilometre Array Observatory, see, e.g., \cite{Klockner:2015rqa,Rocha:2022gog}.}. The best candidates to do so are massive, passively evolving galaxies that formed rapidly at $z\sim 2-3$, over timescales of $\sim0.3$ Gyr. These galaxies exhibit remarkably uniform evolution, meaning their stellar populations are comparable regardless of redshift. This homogeneity allows their age differences to trace the time elapsed between redshifts, effectively mapping the expansion history of the universe.
To extract this information, one relies on stellar population synthesis (SPS) models and their spectral energy distributions. While redshifts can be accurately determined through spectral line analysis, ages are not direct observables and techniques such as photometry or single spectral regions (e.g., the D4000 break \citep{Moresco:2020fbm}) have to be adopted.

In this work, we use a compilation of 33 measurements of $H(z)$ obtained with the CCH method in the redshift range $0.07<z<1.965$. We refer the reader to Table I of \cite{Favale:2023lnp} (and references therein), which we update here with the addition of the point at $z=1.26$ from \cite{Tomasetti:2023kek}. Existing correlations between some of these points have been duly taken into account through the covariance matrix, $C_{\rm{CCH}}$, built upon recipes in \cite{Moresco:2020fbm} and computed as \footnote{\url{https://gitlab.com/mmoresco/CCcovariance}}

\begin{equation}\label{eq:covM}
C_{\rm{CCH},ij}=C_{ij}^{stat}+C_{ij}^{sys}\,.
\end{equation}%
In Sec. \ref{sec:cch_fore}, we will comment on the systematic contributions, $C_{ij}^{sys}$, affecting the CCH method.

\subsubsection{SNIa from DESY5}\label{sec:snIa_curr}
The recent SNIa compilation released by the DES collaboration \citep{DES:2024jxu} consists of 1829 objects. While being the largest single sample survey to date (1635 DES SNIa), it covers the redshift range $0.025\leq z\leq1.12$. At low redshifts, $z<0.1$, the data set is complemented by 194 spectroscopically confirmed low-$z$ SNIa from external samples, 90\% of which are in common with the Pantheon+ compilation \citep{Scolnic:2021amr}. At higher redshifts ($z>0.1$), the overlap between the two samples consists of 145 DES SNIa events and in particular, for $z>0.5$, the number of high-quality SNIa is $\sim5$ times higher in DESY5 \citep{DES:2024jxu}. Some arguments in favor of a potential bias in the DES SNIa sample have been recently raised by \cite{Efstathiou:2024xcq}, see also  \citep{Gialamas:2024lyw,Notari:2024zmi,Huang:2025som,Ormondroyd:2025phk}. However, the DES collaboration have provided arguments against these concerns \citep{DES:2025tir}\footnote{Shortly before the release of this work, \cite{Popovic:2025qnc} re-calibrated the DESY5 sample, yielding a lower evidence for dark energy dynamics compared to the original DESY5 sample, but still slightly larger than the one found with Pantheon+.}. In view of this discussion, and for the sake of completeness, we also present results obtained with Pantheon+ data\footnote{\url{https://github.com/PantheonPlusSH0ES/DataRelease}}.

As standardizable candles, SNIa are expected to have a constant absolute magnitude $M$ after the standardization, independent of the redshift. Hence, if $M$ is known, this property allows us to infer the luminosity distance of the source, $D_L(z)$, directly from measurements of the apparent magnitude, $m(z)$, via the relation

\begin{equation}\label{eq:m_app}
    m(z) = M + 25 + 5\log_{10}\Big(\frac{D_L(z)}{1 \rm{Mpc}}\Big)\,,
\end{equation}%
where
\begin{equation}\label{eq:DLx}
    D_L(z) = \frac{c}{H_0}(1+z)\chi(z)\,.
\end{equation}%
The luminosity distance depends on the spatial curvature, $\Omega_k$, through the function
\begin{equation}
    \chi(z) \equiv \begin{cases}
      \frac{1}{\sqrt{\Omega_k}}\sinh(\sqrt{\Omega_k}\zeta(z)) \hspace{20pt} {\rm if} \hspace{10pt} \Omega_k >0\\
      \zeta(z) \hspace{78pt} {\rm if}\hspace{10pt} \Omega_k=0\\
      \frac{1}{\sqrt{-\Omega_k}}\sin(\sqrt{-\Omega_k}\zeta(z)) \hspace{13pt} {\rm if} \hspace{10pt} \Omega_k<0\,
    \end{cases}
\end{equation}%
with 
\begin{equation}
    \zeta(z) = \int^{z}_{0} \frac{H_0}{H(z^{\prime})}dz^{\prime}\,.
\end{equation}%
Thus, in a FLRW universe, the luminosity distance of a given object can be calculated via the universe's expansion rate $H(z)$.

In this work, we make use of the DESY5 corrected apparent magnitudes $m(z)$ along with the corresponding covariance matrix, $C_{\rm{DES}}$, which accounts for both the statistical and systematic uncertainties\footnote{\url{https://github.com/des-science/DES-SN5YR}}.

\subsubsection{BAO from DESI}\label{sec:desi_curr}

The luminosity distance in Eq. \eqref{eq:DLx} is related to the angular diameter distance, $D_A(z)$, through the Etherington (or cosmic distance duality) relation as \citep{Etherington:1933}

\begin{equation}\label{eq:D_A}
    D_A(z) = \frac{D_L(z)}{(1+z)^{2}}\,.
\end{equation}%
This expression holds in any metric theory of gravity with a conserved number of photons that propagate along null geodesics\footnote{Recently, the validity of the cosmic distance duality relation has been the subject of several studies. We refer the reader to, e.g., \cite{Euclid:2020ojp,Renzi:2021xii, Favale:2024sdq, Qi:2024acx, Tang:2024zkc, Keil:2025ysb, Teixeira:2025czm} and references therein for some model-independent analyses carried out using the latest low-redshift observations.}. Angular diameter distances can be used to map the expansion history and the geometry of the universe, as they relate the physical size of an object to its observed angular size on the sky. BAO measurements from galaxy surveys rely on $r_d$ as a standard ruler to infer cosmic distances across redshifts. In particular, the BAO signal can be given either in terms of transversal information, providing data on $D_M(z)/r_d$, with $D_M(z) = (1+z)D_A(z)$ the comoving angular diameter distance, or in terms of radial information from $H(z)r_d$.
For redshift bins with low signal-to-noise ratio, only the dilation scale to $r_d$ ratio can be measured, and it is defined as

\begin{equation}\label{eq:DVrd}
    \frac{D_V(z)}{r_d} = \frac{\left[zD_M^2(z)D_H(z)\right]^{1/3}}{r_d}\,,
\end{equation}%
where $D_H(z) = c/H(z)$.
An estimate of the sound horizon $r_d$ can then be used to calibrate these distances.

The DESI collaboration \citep{DESI:2024uvr} recently released anisotropic (3D) BAO data from galaxy, quasar and Lyman forest tracers from the first year (DR1) and three years (DR2) of observations. To match the redshift range of our calibrators (CCH), we only use BAO from the following galaxy tracers: the Bright Galaxy Sample (BGS) in $0.1 < z < 0.4$, the Luminous Red Galaxy Sample (LRG) in $0.4 < z < 0.6$ and $0.6 < z < 0.8$, LRG + Emission Line Galaxy Sample (ELG) in $0.8 < z < 1.1$ and ELG data in $1.1 < z < 1.6$. 
Together, these tracers provide DESI BAO measurements covering the redshift range $0.30 \leq z \leq 1.32$, complemented by the QSO result at $z=1.49$. DESI BAO given in terms of $D_M/r_d$ and $D_H/r_d$ are measured at the same redshift and have therefore some degree of correlation, which we account for through the corresponding covariance matrix $C_{\rm{DESI}}$. For completeness, we perform our analysis using separately DESI-DR1 and DESI-DR2 BAO data, as listed in \cite{DESI:2024uvr} and \cite{DESI:2025zgx}, respectively. 

\subsubsection{Angular BAO}\label{sec:2DBAO}
When the BAO signal is extracted from the angular position of the BAO peak, the analysis is carried out in the two-dimensional parameter space by making use of the two-point angular correlation function or the angular power spectrum; therefore, any information about the radial BAO scale and thus $H(z)$ is largely lost. The provided information is given in terms of the angular (transversal) scale $\theta(z)=r_d/D_M(z)$. In this case, a fiducial cosmology to convert measured redshift and angles into comoving distances is not required. For this reason, 2D measurements are claimed to be less affected by model dependencies than the standard 3D approach \citep{Alcaniz:2016ryy, deCarvalho:2017xye, deCarvalho:2021azj}. To study whether and how our results are affected by the adoption of 2D BAO measurements, we make use of 16 angular BAO data points on $\theta(z)$ in $0.11\leq z \leq 2.225$ \citep{Carvalho:2015ica, Carvalho:2017tuu,Alcaniz:2016ryy, deCarvalho:2017xye,deCarvalho:2021azj, DES:2024pwq}, collected in Table 2 of \cite{Favale:2024sdq}.

\subsection{Forecast data from ongoing and future surveys}\label{subsec:fore}
In this section, we detail the mock data sets that we build to perform forecasts for ongoing and future surveys. For all observables of our interest -- namely $H(z)$, $m(z)$ and the BAO distances -- we adopt as fiducial data vector the corresponding flat $\Lambda$CDM prediction obtained using the {\it Planck} TT,TE,EE+lowE+lensing best-fit parameters from \cite{Aghanim:2018eyx}, and $M=-19.4$. Table \ref{tab:mock_data} summarizes the specifications for each mock data set to guide the reader through the different scenarios.

\subsubsection{CCH from Euclid, WST and ATLAS}\label{sec:cch_fore}

We generate data on $H(z)$ following the recipes for the error budget and the redshift range provided in \cite{Moresco:2022phi,Moresco:2024wmr}. Specifically, the redshift range is adopted from \cite{Moresco:2024wmr}, which selects 20 points at low-redshifts ($z<1.5$) that will be provided by future surveys, such as the Wide-field Spectroscopic Telescope (WST, \cite{WST:2024rai}) or the Astrophysics Telescope for Large Area Spectroscopy Probe (ATLAS, \cite{Wang:2019jig}), along with 5 high-redshift points ($1.5<z<2$), from the Euclid mission, for a total of 25 data points\footnote{To be consistent across all the forecasts presented for SNIa, BAO and CCH data, we choose not to combine current and future CCH data in this work. For cosmological constraints derived from such a combination, we refer the reader to \citep{Moresco:2022phi, Moresco:2024wmr}. For instance, in \cite{Moresco:2022phi}, it has been shown that fitting an open $w$CDM model with future low-$z$ CCH already reduces the $H_{0}$ uncertainty to $\approx 4.4\%$, and the combination with current data only marginally improves it, to $\approx 4\%$.}. A statistical uncertainty of 5\% is assumed for measurements of $H(z)$ throughout the redshift range, as expected for WST over $z < 1.5$ \citep{WST:2024rai}, and for Euclid. The ESA mission is likely to provide up to a few thousand passively evolving galaxies at $1.5<z<2$, increasing the current statistical sample by two orders of magnitude \citep{Moresco:2022phi}. This would enable measurements of the Hubble function to improve upon the current precision, which ranges between 10\% and 25\% \citep{Tomasetti:2023kek}.
 
We base our treatment of systematic uncertainties that will enter Eq. \eqref{eq:covM} on the recipes provided in \cite{Moresco:2022phi}.

When massive and passively evolving galaxies are employed as CCH candidates, their intrinsic physical properties introduce diagonal errors, given that these uncertainties are uncorrelated between galaxies at different redshifts. This is the case of, e.g., possible contamination by younger stars which causes the measurement of younger ages, and stellar metallicity, currently resulting in a $<0.5\%$ and 4\% level error on $H(z)$, respectively. However, this is bound to be kept under control with high spectral resolution and high-S/N data \citep{Moresco:2020fbm}, as those considered in this forecast. For this reason, we regard them as negligible. The same applies to contributions from progenitor bias and mass dependence, which are already considered negligible and therefore not accounted for here.
We follow the prescriptions used in \cite{Moresco:2022phi} and \cite{Moresco:2024wmr}, in which the systematic contributions include the impact of the initial mass function (IMF) and SPS models (we call this \textit{pessimistic scenario}). The uncertainty in $H(z)$ due to the IMF is estimated at 0.4\%. The effect of considering different SPS models impacts the $H(z)$ measurement with a 9\% error on average, with a decreasing trend with increasing redshift in $0<z<1.5$. This different trend is related to the behaviour of the slopes of the D4000n-age relations for the different models \citep{Moresco:2020fbm}. By excluding the most discrepant SPS model, the corresponding contribution can be reduced to $\sim4.5\%$. Nevertheless, the choice of SPS model remains the dominant source of uncertainty in the total error budget, and these errors are highly correlated across different spectra.

As suggested in \cite{Moresco:2022phi} and \cite{Moresco:2024wmr}, we also study an \textit{optimistic scenario} which minimizes the impact of the SPS models. In this case, the associated systematic uncertainty is assumed to be resolved, thanks to advancements in future spectroscopic analyses and observations that are expected to aid in identifying and narrowing down the set of SPS models that best reproduce the data. As a result, the IMF-related uncertainty is the only systematic contribution considered in this scenario.

The covariance matrix in Eq. \eqref{eq:covM} will be

\begin{equation}\label{eq:covMfore}
    C_{\rm{CCH^{*}},ij}=C_{ij}^{stat}+C_{ij}^{sys} = C_{ij}^{stat} + C_{ij}^{IMF} + C_{ij}^{SPS}\,,
    \end{equation}%
where, following the previous discussion, the presence of the last term depends on the scenario considered (pessimistic vs optimistic). Hereafter, we will call this mock data set \textbf{CCH$^{*}$}.

\subsubsection{SNIa from LSST}\label{sec:snIa_fore}

LSST will play a pivotal role not only in advancing astrophysical research, but also in driving significant progress in cosmological analyses. Already with an early data release corresponding to roughly half of its final mission target, LSST is expected to be able to constrain a constant dark energy equation of state $w$ at the sub-5\% level and to obtain a precision of 0.05 on $w_0$ for a time-varying $w_0w_a$CDM equation of state \citep{2009lsst}.
The survey is projected to increase the SNIa sample size by up to a factor of 100 compared to previous samples, with redshifts up to $z\sim1.2$ \citep{LSSTDarkEnergyScience:2021laz}. It will use a single instrument, which may help mitigate calibration issues present in samples built from different telescopes. A good-enough photometry will enable sufficiently well-sampled and accurate light curves and host galaxy redshifts, whose distribution mean will be around $z\sim0.4$ \citep{LSSTDarkEnergyScience:2012kar}.

According to this framework, we proceed with the SNIa forecast as outlined below. The redshift distribution of the final LSST sample is assumed to be comparable to DESY5 \citep{LSSTDarkEnergyScience:2021laz}; however, the LSST sample size will be substantially larger. To account for this, we treat the DESY5 data set as a binned proxy for the final-year LSST data. We rescale the statistical component of the DESY5 covariance matrix by a factor $n=N_{\rm{DES}}/N_{\rm{LSST}}$, with $N_{\rm{DES}} = 1829$ and $N_{\rm{LSST}}=10^{5}$. This number corresponds to the predicted 10-year final sample size (hereafter \textbf{LSST-Y10}), which results in a rescaled statistical error $\sigma^{stat}_{\rm{LSST-Y10}} \approx \sigma^{stat}_{\rm{DES}} / 7 $. The increase in sample size $N$ affects only the statistical component of the total uncertainty, which scales as $1/\sqrt{N}$. Reducing the statistical uncertainty enhances the relative contribution of systematics to the overall error budget (see \cite{Kim:2003mq}). For the systematic component of the covariance matrix, we keep the estimate from DESY5. This is a conservative choice, as control of systematic uncertainties may improve with the LSST sample. For this reason, we opt to present two distinct scenarios. The one previously discussed will be our conservative approach, and is referred to as the \textit{pessimistic scenario}. In contrast, we also consider an \textit{optimistic scenario}, in which all systematic uncertainties are assumed to be negligible\footnote{As we will see in Sec. \ref{sec:res_fore}, the impact of the DESY5 correlations is derisory when SNIa are combined with CCH in forecast analyses. For this reason, we choose not to analyze an intermediate case for the treatment of systematics in the LSST covariance matrix, as the results would remain unchanged.}.

\begin{table*}
\centering
\caption{Summary description of the mock data sets built for each probe (CCH, SNIa and BAO) and used in the forecast analysis presented in Sec. \ref{sec:res_fore}. For further details on the samples, see Sec. \ref{subsec:fore}.}\label{tab:mock_data}
\resizebox{\textwidth}{!}{%
\begin{tabular}{lccc}
Data set & Surveys & Sample size & Description \\ \hline\hline
CCH$^{*}_{\rm pes}$ & WST, ATLAS ($z<1.5$); Euclid ($1.5<z<2$) & 25 & Systematics include IMF and SPS model contributions\\ \hline
CCH$^{*}_{\rm opt}$ & - & -& Systematics include only IMF contribution\\ \hline
LSST-Y10$_{\rm pes}$ & LSST year 10 ($z<1.2$)& 1829 (binned proxy for $10^{5}$) & DESY5 systematics; statistical errors rescaled by $1/\sqrt{N_{\rm{DES}}/N_{\rm{LSST}}}$\\ \hline
LSST-Y10$_{\rm opt}$ & - & -& Systematics assumed negligible\\ \hline
DESI$^{*}$ & DESI year 5 ($0.05\leq z \leq2.05$) & 42 & $D_A/r_d$ and $H(z)r_d$ data and systematics from \cite{DESI:2023dwi} \\
\hline
\end{tabular}
}
\end{table*}

\subsubsection{BAO from DESI}\label{sec:bao_fore}
In \cite{DESI:2023dwi}, the DESI collaboration reported BAO and redshift space distortions forecasts that represent the expected final target of the mission. They account for a predicted full footprint of 14,000 deg2 for each tracer. This area is used for cosmological forecasts, which set DESI as a Stage-IV dark energy experiment. Indeed, the precision on the isotropic BAO distance scale from BGS and LRG at redshifts $z < 1.1$ is expected to be of 0.28\%, while the ELG and quasar sample in $ 1.1 < z < 1.9$ will provide BAO measurements at 0.37\% precision. These regions represent the redshift range of our interest, given the last CCH point at $z=1.9$. However, CCH forecasts reach redshift $\sim2$ and in such a case, the BAO feature measured in the Ly-$\alpha$ forest at higher redshift ($z=2.05$) is also included in the analysis.
In particular, we employ the predicted $D_A/r_d$ and $H(z)r_d$ measurements provided in Table 7 of \cite{DESI:2023dwi}, for a total of 42 points between $0.05\leq z \leq2.05$. Predicted redshifts and the corresponding covariance matrix that we use are available as supplementary materials \footnote{\url{https://doi.org/10.5281/zenodo.10063934}}. We denote this data set as \textbf{DESI$^{*}$}.

\section{Methodology}\label{sec:method}
This work adopts the methodology outlined in \cite{Favale:2023lnp}, whose main steps are summarised below. These steps are applied to both the current data analysis and the forecast analysis. For further methodological details, readers are referred to the aforementioned reference.

\subsection{Statistical tool}\label{sec:method_2}
In this work, we aim to measure the calibrators $M$ and $r_d$, as well as to estimate the curvature parameter, by using CCH data as a probe of the expansion history of the universe (see Sec. \ref{sec:cch_curr}). Thus, we need to start from agnostic reconstructions of $H(z)$.

We leverage the Gaussian Process technique to enable data-driven reconstruction of the functions of interest, with minimal assumptions \citep{2006gpml.book.....R}. The GP can be regarded as a generalization of a multivariate Gaussian, defined as $f(z) \sim GP(\mu(z), D(z, \tilde{z}))$, where $\mu(z)$ the mean function and $D(z, \tilde{z})$ its covariance matrix. This matrix is a combination of the covariance matrix of the data, $C(z, \tilde{z})$ (being $z,\tilde{z}$ the locations of the $Y$ data values) and of a kernel function, $K(z, \tilde{z})$. The choice of the kernel falls within the minimal assumption above-mentioned, as it controls the shape of the reconstructed function, $f^{\star}$. The latter is defined by

\begin{equation}\label{eq:GPmean}
    \bar f^{\star}=\mu^{\star}+ K(z^{\star},z)[K(z,z)+C(z,z)]^{-1}(Y-\mu)\,,
\end{equation}
with the associated covariance 

\begin{equation}\label{eq:GPcov}
   {\rm cov}(f^{\star})= K(z^{\star},z^{\star})-K(z^{\star},z)[K(z,z)+C]^{-1}K(z,z^{\star})\,,
\end{equation}
where $\mu^{\star}\equiv\mu(z^\star)$ is the a priori mean of the reconstructed function at $z^\star$.
The kernel provides a functional form for the assumed correlations between points, even in regions where data are not available. A set of hyperparameters controls the strength of the fluctuations and the correlation length between two separate points, and they can be determined by the optimization (maximization) or marginalization (sampling) of the GP likelihood.
Indeed, within a Bayesian approach, the full distribution of the kernel hyperparameters should be obtained in order to account for correlations between them and propagate their uncertainties to the final reconstructed function $f^{\star}$ (see, e.g., \cite{Shafieloo:2012, Gomez-Valent:2018hwc, Hwang:2022hla, Johnson:2025blf, Ruchika:2025mkx}).

Following this, we use the public package {\it Gaussian Processes in Python} (\texttt{GaPP}) \footnote{\url{https://github.com/carlosandrepaes/GaPP}}, developed by \cite{Seikel:2012uu}, to reconstruct $H(z)$ from CCH, at the respective redshift locations of the data points at play.
The robustness of these reconstructions is supported by and validated in our previous work \citep{Favale:2023lnp}, which thoroughly examines the effects of kernel selection, prior mean function and hyperparameter marginalization. See the quoted reference for a comprehensive discussion.
For completeness, and in light of new data considered here, we repeat a kernel-sensitivity test in the present work. The results are reported in Appendix \ref{app:kernel}.

We here emphasize that the GP posterior mean function in Eq. \eqref{eq:GPmean} explicitly depends on the data uncertainties. For a zero prior mean function ($\mu=0$), the expectation value of the posterior mean at the data location ($z^\star=z$) generally differs from the input data vector $Y$, except in the limit of vanishing noise. For $\mu\ne 0$, instead, the only way to force the proposed mean function to pass exactly through the data points would be to set $\mu=Y$, so that $f^{\star}\equiv\mu(z^\star)$. However, in realistic applications, the true underlying function is unknown, and such a choice is not possible. Moreover, our method is designed to remain model-agnostic: by setting $\mu=0$, we assume no prior knowledge of the functional form behind the data. As a result, an offset might arise as an intrinsic feature of the GP reconstruction (see also \citealt{Seikel:2012uu,Perenon:2021uom, Perenon:2022fgw}). In this work, we explicitly quantify this bias in order to assess its magnitude and potential impact on our results (see Sec. \ref{sec:method_3}).

\subsection{Joint analyses: a grid-search approach}\label{sec:method_3}
We employ the GP-reconstructed functions in a grid-search method. First, CCH and SNIa data sets are used to obtain joint constraints in the plane $(M,\Omega_k)$. The same for CCH and BAO in the plane $(\Omega_k, r_d)$. These two-dimensional joint analyses are then combined to exploit the full three-dimensional parameter space of $(M,\Omega_k,r_d)$. We remark here that in our analyses we constrain $\Omega_k$ through its effect on the geometrical pre-factor of the luminosity distance and not on $H(z)$, since we reconstruct the Hubble function directly with GP+CCH. 
The size and resolution of each grid can vary depending on the data set employed. We always assume flat priors for $M$, $\Omega_k$, $r_d$, much wider than the uncertainties obtained in our analyses.

For each point of the grid, which is characterized by the pair $(M,\Omega_k)$ or $(\Omega_k, r_d)$, we generate $N$ realizations of $H(z)$, which are then employed to obtain $N$ realizations containing the values of the luminosity distances (Eq. \ref{eq:DLx}) at the SNIa redshifts, or angular diameter distances and dilation scales (Eqs. \eqref{eq:D_A}, \eqref{eq:DVrd}) and values of $H(z)$ at the BAO redshifts. These $N$ realizations are easily converted into $N$ realizations of reconstructed SNIa and BAO observables at the corresponding redshifts. This information is then used to perform a $\chi^2$ analysis using the SNIa or BAO data.
For each realization $I=1,..,N$ in the $\mu$-th knot, we have
 
\begin{equation}\label{eq:chi2_mu}
    \chi_{X, \mu,I}^{2} = \sum_{k,l=1}^{D_X} [f_k - f^{rec}_{\mu,I,k}]( C_X^{-1})_{kl}[ f_l - f^{rec}_{\mu,I,l} ]\,,
\end{equation}
where $C_{X}$ is here the covariance matrix of the data set $X$ employed -- either SNIa or BAO --, and $D_X$ the corresponding total number of data points. Notice that each index $k,l$ labels an individual data point in the data set $X$, as the latter can include multiple measurements at the same redshift $z$.
Each knot of the grid is weighted by

\begin{equation}\label{eq:wmu}
w_\mu\propto B_\mu \sum_{I=1}^{N} \exp(-\chi_{X,\mu,I}^2/2)\,,
\end{equation}
where the prefactor $B_{\mu}$ accounts for the bins size at the $\mu$-th knot.
If we denote with $P_{AB}$ the two-dimensional probability for the parameters $A$ and $B$, then its evaluation will be as follows

\begin{equation}
P_{AB}(a,b)=\frac{w_{\mu\to (a,b) }}{\sum\limits_{\beta} w_\beta}\,.
\end{equation}
While the denominator accounts for contributions from all knots, the numerator is restricted to the knot corresponding to the parameter values $a$ and $b$ of $A$ and $B$, respectively. 
Analogously, the one-dimensional posterior probability for each parameter $A$, $P_A$, is

\begin{equation}\label{eq:P_chi2eff}
    P_A(a)=\frac{\sum\limits_{\mu\to a} w_\mu}{\sum\limits_{\beta} w_\beta}\,.
\end{equation}
The sum in the numerator is now limited to the knots associated with the value $a$ of $A$. Best-fit values for each parameter are therefore obtained by maximizing Eq. \eqref{eq:wmu}.

Notice that, as mentioned before, the 2D analyses are combined together to explore the full parameter space. From Eq. \eqref{eq:chi2_mu}, this simply translates in computing $\chi^{2}_{X\Tilde{X}} = \chi^{2}_{X} + \chi^{2}_{\Tilde{X}}$, as the BAO and SNIa data are independent.

We follow this approach both when employing the latest data sets and for the forecasts, whose results are shown, respectively, in Sec. \ref{sec:res_curr} and Sec. \ref{sec:res_fore}.

As noticed in Sec. \ref{sec:method_2}, an intrinsic offset in the GP reconstructions is expected by construction of the posterior mean function, particularly when adopting a zero prior mean. This offset propagates through the analysis and can ultimately induce a bias on the inferred parameters ($M, \Omega_k, r_d$). We therefore find it necessary to quantify this effect using mock data, for which the fiducial values of the parameters are known. Indeed, within our analysis, this bias on parameters may arise from deviations in the GP reconstruction of $H(z)$ with respect to the fiducial input. The GP has access only to a finite set of data points which, in the case of the mock data, correspond to the values of the fiducial $H(z)$ curve evaluated at the redshifts of the simulated observations. In addition, the data points are provided with a covariance matrix that describes their expected uncertainties. As a consequence, the reconstructed posterior mean Eq.~\eqref{eq:GPmean} does not match exactly the true underlying function, even for mock data. Each realization drawn from the GP posterior is distributed around this posterior mean, and averaging over many realizations simply recovers the same mean. Consequently, the bias embedded in the posterior mean is preserved and propagates to the final reconstructed quantities. We estimate this effect as the difference between the recovered and the fiducial value of each parameter $A$, $\tilde{b} = A_{\rm rec} - A_{\rm fid}$. Accounting for the sign of $\tilde{b}$ is essential, as the direction of the offset is physically informative: a systematic shift, for instance towards higher values of $H(z)$, implies a smaller sound horizon and higher absolute magnitude. The sign of $\tilde{b}$ thus verifies that the parameter recovery aligns with the expected physical trend. Its estimated values are reported in Table \ref{tab:res_curr} for one representative combination of current data, and displayed in the figures presenting the forecast results (see Sec. \ref{sec:res}).

An additional step is introduced specifically for the forecast analysis. Indeed, given the improved accuracy of the samples considered in this study, we expect that the grid resolution and the number of $N$ realizations needed to obtain well-sampled posteriors for the parameters of interest increase significantly. The effect is even more pronounced when considering the \textit{mixed} cases, which are aimed at illustrating the influence of improvement in a single data set (see Sec. \ref{sec:res_fore}). For instance, the analysis with LSST-Y10+CCH reveals a large difference between the current CCH error bars and the improved statistics achieved by the future SNIa sample. Current CCH measurements exhibit relative uncertainties of 7–60\% across the probed redshift range, in contrast to the projected precision of 0.1–0.3\% attainable with LSST-Y10. This difference translates into a high computational cost to achieve convergence. For this reason, we opt to fit the mildly noisy posteriors to obtain smoother curves. Thus, accurate constraints on the parameters of interest can still be extracted while preserving the overall shape and width of the original sampled distributions. We refer the reader to Appendix \ref{app:fit} for details on the approach followed. Our main results are displayed and discussed in Sec. \ref{sec:res}.

\subsubsection{An estimate of $H_0$ from the CCH-calibrated ladders}\label{sec:H0}

As the Hubble tension continues to gain significance and fuel active debate, it is crucial to assess where model-independent estimates of $H_0$ stand in the present (and future) landscape. For this reason, as a final step of the analysis, we estimate which is the value of the Hubble parameter preferred by the CCH ladder calibration. To do so, as done within the context of the local distance ladder analyses, we apply the following cosmographical expansion for the luminosity distance\footnote{We can neglect here curvature corrections as they are of third order in $z$, therefore playing a negligible role within the Hubble flow.},

\begin{equation}\label{eq:H0}
D_L(z)=czH^{-1}_0\left[1+\frac{z}{2}\left(1-q_0\right)\right]+\mathcal{O}(z^3)\,,
\end{equation}
with $q_0$ the deceleration parameter.
The luminosity distance is computed using the SNIa apparent magnitudes in the Hubble flow ($0.023 <z<0.15$) and the constraint on $M$ obtained from our final 3D joint analysis.
This is analogous to the analysis performed by the SH0ES collaboration, which uses the Cepheid-calibrated value of $M$ -- see e.g. \cite{Riess:2016jrr, Riess:2021jrx} -- and a fixed value for the deceleration parameter, $q_0 =-0.55$. At very low redshifts ($z\lesssim0.1$), the impact of the first-order correction in $q_0$ is small, since $z^{2}\ll1$. Although this makes $q_0$ estimation challenging from low-redshift data alone (especially if one has a large uncertainty on $M$ that dominates the error budget), we opt to fit both parameters simultaneously. This way, uncertainties in $q_0$ can be propagated into uncertainties (or potential biases) on the estimation of $H_0$. We discuss the corresponding results in Sec. \ref{sec:res}.

\section{Results and Discussion}\label{sec:res}
This section is devoted to the presentation and discussion of the results of the analysis described in Sec. \ref{sec:method_3}.
First, in Sec. \ref{sec:res_curr}, we apply the methodology with the current available data whereas in Sec. \ref{sec:res_fore} we present the results of the forecast analysis.

\subsection{Analysis with current data}\label{sec:res_curr}

\begin{table*}
\centering
\caption{Constraints at 68\% C.L. on the ladder calibrators, $M$ and $r_d$, and on the curvature parameter $\Omega_k$, from the joint analyses presented in Sec. \ref{sec:res_curr}. In the last two columns, we report the cosmographical constraints on $H_0$ and $q_0$, cf. Eq. \eqref{eq:H0}, while the last two rows include the results in which we test the impact of changing SNIa data (Pantheon+) or the type of BAO (2D, angular). For the analysis combining CCH, DESY5 and DESI DR2, we additionally report the result on $H_0$ by fixing $q_0 = -0.55$, and we show (in parentheses) the methodological bias $\tilde{b}$ estimated from mock data using fiducial input parameters. See main text in Secs. \ref{sec:method_3} and \ref{sec:res_curr} for details.}\label{tab:res_curr}
\begin{tabular}{lrrrrr}
&$M$ & $\Omega_k$ & $r_d$ [Mpc] & $H_0$ [km/s/Mpc] & $q_0$ \\ \hline\hline
CCH+DESY5 &$-19.277_{-0.106}^{+0.104}$ ($+0.017$) & $-0.48_{-0.27}^{+0.32}$ ($-0.002$)& - & - & -\\
CCH+DESI DR1 & - & $-0.03_{-0.16}^{+0.17}$ & $146.10_{-5.18}^{+5.38}$ & - & -\\
CCH+DESI DR2  & - & $-0.12\pm0.09$ ($-0.007$)& $144.20_{-4.88}^{+5.18}$ ($-0.57$)& - & -\\
CCH+DESY5+DESI DR1 & $-19.330_{-0.099}^{+0.095}$& $-0.113_{-0.155}^{+0.150}$ & $145.30_{-5.08}^{+5.28}$ & $68.60_{-3.30}^{+2.90}$ & $-0.06_{-0.33}^{+0.34}$\\
CCH+DESY5+DESI DR2  & $-19.324_{-0.095}^{+0.092}$ ($+0.029$)& $-0.143\pm0.085$ ($-0.007$)& $144.00_{-4.88}^{+5.08}$ ($-0.57$)& $68.83_{-3.07}^{+3.03}$ & $-0.05_{-0.32}^{+0.34}$\\
& & & & $69.60_{-3.01}^{+3.00}$ & $-0.55$ (fixed)\\
\hline
CCH+PAN+DESI DR2 & $-19.309\pm0.088$ & $-0.107_{-0.083}^{+0.079}$ & $144.00_{-4.68}^{+5.38}$ & $72.12_{-2.95}^{+2.97}$ & $-0.45\pm0.19$\\
\hline
CCH+DESY5+2D\_BAO & $-19.294_{-0.106}^{+0.101}$ & $-0.362_{-0.240}^{+0.259}$ & $147.95_{-6.84}^{+7.32}$ & $69.85_{-3.58}^{+3.19}$ & $-0.04_{-0.31}^{+0.34}$\\
\hline
\end{tabular}
\end{table*}

\begin{figure}
    \centering    \includegraphics[width=\linewidth]{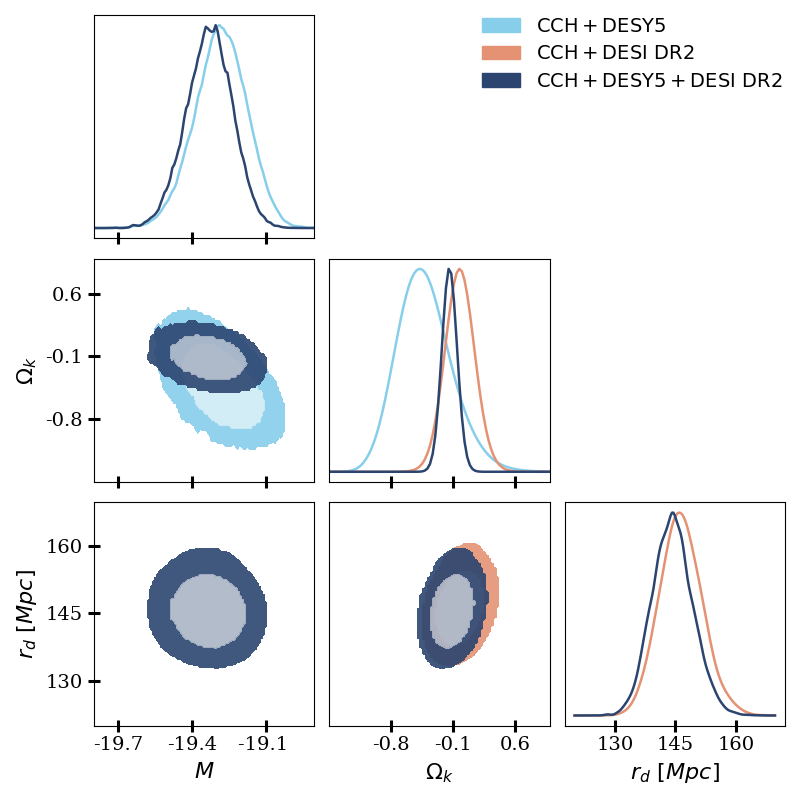}
    \caption{Two-dimensional contour plots in all the planes of the parameter space $(M, \Omega_k, r_d)$ and the corresponding one-dimensional posterior probability distributions obtained from the analyses described in Sec. \ref{sec:res_curr}. The 2D contours are evaluated at 68\% and 95\% C.L.}
    \label{fig:triangleDR2}
\end{figure}

In Table \ref{tab:res_curr} we present the constraints at 68\% C.L. on $M$, $\Omega_k$, $r_d$, $H_0$ and $q_0$ obtained with the most recent data from current surveys. The one-dimensional posteriors and the confidence regions at 68\% and 95\% C.L. in all the planes of the parameter space for $M$, $\Omega_k$ and $r_d$ are shown in Fig. \ref{fig:triangleDR2}, in particular for the combination of CCH with DESY5 and DESI DR2 data.

The CCH+DESY5 analysis constrains the SNIa absolute magnitude as $M=-19.277_{-0.106}^{+0.104}$ and the curvature parameter as $\Omega_k=-0.48_{-0.27}^{+0.33}$. In \cite{Favale:2023lnp}, using CCH data and SNIa from Pantheon+ (hereafter PAN), we obtained $M=-19.344^{+0.116}_{-0.090}$ and $\Omega_k=-0.07^{+0.27}_{-0.21}$.
The new constraint for $M$ remains quite stable. Its uncertainty is driven by the errors on $m(z)$ and $\log D_L(z)$ (see Eq. \eqref{eq:m_app}). However, the dominant contribution arises from the luminosity distance, as it is the quantity derived from the GP reconstruction with CCH, which are not as precise as SNIa measurements. Since the CCH data set is the same between the analyses performed with DESY5 and Pantheon+, the determination of $M$ appears largely unaffected.
On the other hand, $\Omega_k$ from Pantheon+ is more precise than that obtained here with DESY5, and this can be attributed mainly to the following reason. $\Omega_k$ is more sensitive to higher redshifts. The Pantheon+ SNIa span the redshift range $[0.001,2.26]$, compared to the region $[0.025,1.12]$ of DESY5. As a result, the current 2D analysis yields relatively weaker constraints on the curvature parameter. We have checked that, if we cut the Pantheon+ sample at $z<1.12$, i.e., taking the largest redshift as in DESY5, and perform the CCH+PAN analysis, we indeed obtain a 40\% increase in the uncertainty of $\Omega_k$ compared to the one obtained exploiting all the SNIa up to $z=2.26$, hence obtaining an uncertainty closer to that of DESY5. It should be noted that the broader redshift coverage of Pantheon+ compensates for its larger distance errors when these SNIa are used to obtain cosmological constraints, whose uncertainties are smaller than those obtained using DESY5, also because the latter include additional systematic uncertainties \citep{DES:2025tir}.

The DESY5 mild preference at 1.6$\sigma$ for a negative $\Omega_k$ is erased by DESI DR1 data in the CCH+DESI DR1 analysis, while the most stringent result with DESI DR2 partially recovers it at 1.3$\sigma$. The second data release improved upon the constraining power of the BAO measurements from DR1; therefore, the most precise results are expected from the former. Nonetheless, we find it valuable to assess the effect of both data releases in this work. The impact of DESI DR2 on $\Omega_k$ is substantial, reducing its uncertainty by a factor $\sim$2. On the other hand, the constraint on $r_d$ decreases only by 4\%.

The 3D joint analysis performed with DESI DR2 leads to $M = -19.324_{-0.095}^{+0.092}$, $\Omega_k = -0.143\pm0.085$ and $r_d = (144.00_{-4.88}^{+5.08})$ Mpc. Again, we can compare these updated constraints with those obtained in \cite{Favale:2023lnp} with SNIa from Pantheon+ and BAO from several galaxy surveys (6dFGS, BOSS, eBOSS, WiggleZ and DES Y3), which read: $M=-19.314^{+0.086}_{-0.108}$, $\Omega_k=-0.07^{+0.12}_{-0.15}$ and $r_d=(142.3\pm 5.3)$ Mpc. The addition of DESI DR2 data clearly balances the increase in the error bars given by DESY5 SNIa, which we discussed above. This makes the updated constraints with current late-time data tighter than those obtained in the previous analysis. The value of the Hubble constant stands at $H_0=(68.83_{-3.07}^{+3.03})$ km/s/Mpc using the value of $M$ obtained within the CCH+DESY5+DESI DR2 analysis. In \cite{Favale:2023lnp}, we obtained $H_0=(71.5\pm 3.1)$ km/s/Mpc. While the error does not move significantly, the central value points to the ballpark of $H_0$ values obtained with CMB probes under the assumption of $\Lambda$CDM. We deem it interesting to study what happens if we still use the DESY5 SNIa in the Hubble flow but considering the SH0ES prior on $M$ obtained from the SH0ES collaboration via the local distance ladder method, $M^{R22}=-19.253\pm0.027$ \citep{Riess:2021jrx}. Remarkably, this leads to $H_0 = (70.84\pm1.16)$ km/s/Mpc. This value is compatible ($<1\sigma$) with those obtained with CCH+DESY5+DESI DR1/DR2 (see Table \ref{tab:res_curr}) and, above all, it is lower than that obtained with the local distance ladder using SNIa from Pantheon+, $H_0=(73.04\pm 1.04)$ km/s/Mpc \citep{Riess:2021jrx}, bringing the tension with the {\it Planck} result, $H_0=(67.36\pm 0.54)$ km/s/Mpc \citep{Aghanim:2018eyx}, below 3$\sigma$. This is true even considering the tightest CMB-based constraint on $H_0$ obtained from the combination of {\it Planck}+ACT+SPT, $H_0 = (67.24\pm0.35)$ km/s/Mpc \citep{SPT-3G:2025bzu}. It is important to highlight this result since, as already mentioned in Sec. \ref{sec:snIa_curr}, there are several ongoing debates on systematics affecting the low-$z$ SNIa of Pantheon+ or DESY5 \citep{Perivolaropoulos:2023iqj, Efstathiou:2024xcq, Notari:2024zmi, DES:2025tir, Huang:2025som}. The vast majority of the SNIa in the Hubble flow that are common to both samples exhibit larger apparent magnitudes in DESY5. These DESY5 SNIa therefore appear fainter, implying larger luminosity distances and, consequently, a lower inferred expansion rate (see Fig. \ref{fig:HF} in Appendix \ref{app:HF}). We note that a very recent re-analysis of the DESY5 sample has been presented in \cite{Popovic:2025qnc}, which we mention here for completeness, and which further motivates careful consideration of SNIa samples. Given these discrepancies, we also study the case in which the Hubble flow of Pantheon+ is used together with the value of $M$ obtained in the CCH+PAN+DESI DR2 analysis, $M=-19.309\pm0.088$. Again, we observe that Pantheon+ prefers a higher value of $H_0$, but with the smallest uncertainty, being $H_0=(72.12_{-2.95}^{+2.97})$ km/s/Mpc, for the reasons explained throughout this section.

Replacing DESI with angular BAO does not significantly change our conclusions regarding the ladder calibrators, as the larger uncertainties prevent drawing robust inferences. This is expected, since these data, being subject to minimal model assumptions, are less precise than the 3D BAO measurements. Consequently, the CCH+DESY5+2D\_BAO combination is the least constraining, cf. Table \ref{tab:res_curr}. For instance, while 2D BAO slightly favors a more closed universe than DESI BAO, with $\Omega_k = -0.362_{-0.240}^{+0.259}$, the uncertainty remains too large for a definitive conclusion. 2D BAO prefer a slightly higher value of the Hubble parameter, $H_0 = (69.85_{-3.58}^{+3.19})$ km/s/Mpc, than the result obtained with anisotropic BAO using the same SNIa sample (DESY5). This is consistent with previous findings in the literature. Indeed, it is already known that angular BAO show a better agreement with the SH0ES estimate \citep{Camarena:2019rmj, Bernui:2023byc, Akarsu:2023mfb, Gomez-Valent:2023uof,Dwivedi:2024okk, Gomez-Valent:2024tdb, Gomez-Valent:2024ejh}.\footnote{Because 2D BAO angular diameter distances are smaller than expected within the standard model, these data tend to favor higher values of the Hubble function, bringing them into closer alignment with the SH0ES measurement.}

As expected, the resulting constraints on the deceleration parameter $q_0$ are very loose for all the cases under study, with the exception of CCH+PAN+DESI DR2 analysis. This is driven by the smaller uncertainty on the corresponding $M$ but especially by the larger number of Pantheon+ SNIa in the Hubble flow, which is more than twice that of DESY5. Our $q_0$ obtained with Pantheon+, $q_0=-0.45\pm 0.19$, is fully compatible with the value employed by SH0ES to estimate $H_0$, $q_0 = -0.55$ \citep{Riess:2016jrr,Riess:2021jrx}\footnote{As stated in \cite{Riess:2021jrx}, the choice of $q_0 = -0.55$ relies on the fact that it has historically provided a good fit to high-redshift SNIa and matches the expectation from the flat $\Lambda$CDM model with $\Omega_m = 0.3$ and $\Omega_\Lambda = 0.7$. Independent works find consistent results, see, e.g., \citep{Haridasu:2018gqm,Gomez-Valent:2018gvm}.}. With DESY5, instead, we obtain $q_0=-0.05^{+0.34}_{-0.32}$. In this case the lack of a strong low-$z$ anchor leaves the fit poorly constrained, compatible with a decelerated universe, but still in full agreement with the value obtained from the model-agnostic reconstruction of $q(z)$ using CMB data from \textit{Planck}, BAO from DESI DR2 and the SNIa from DESY5, $q_0=-0.28^{+0.09}_{-0.08}$ \citep{Gonzalez-Fuentes:2025lei} -- see also \citep{DESI:2025fii}. Our results are therefore consistent with previous results in the literature.
As a cross-check, to assess whether the additional freedom allowed by the loose constraint on $q_0$ could be driving the inferred value of $H_0$ toward lower values, we also estimate $H_0$  from the CCH-calibrated DESY5 ladder, i.e. adopting $M = -19.324_{-0.095}^{+0.092}$ from the CCH+DESY5+DESI DR2 analysis together with the DESY5 Hubble flow SNIa, while fixing the deceleration parameter to $q_0=-0.55$.
We obtain $H_0 = 69.60_{-3.01}^{+3.00}$ km/s/Mpc. This result is consistent at the 0.25$\sigma$ level with the value obtained when allowing $q_0$ to vary freely, $H_0 = 68.83_{-3.07}^{+3.03}$ km/s/Mpc, supporting the conclusion that the decrease in $H_0$ observed with DESY5 may be associated with differences between SNIa samples at the level of the Hubble flow, as explained above. See also Appendix \ref{app:HF}.

As a final comment, we note that Table \ref{tab:res_curr} also reports an estimate of the methodological bias $\tilde{b}$ introduced by the GP (see Sec. \ref{sec:method_3}). While this bias pertains to the parameters being analyzed, it is quantified via a separate analysis which makes use of mock data vectors generated assuming the $\Lambda$CDM model and current data uncertainty. We focus on the combination of CCH, DESY5 and DESI DR2 as a representative test case. Our results show that the $|\tilde{b}|$ values for all parameters are well below the 1$\sigma$ uncertainty on the parameter, being generally $\lesssim0.1\sigma$ and at most $0.3\sigma$ for $M$. For the 3D analysis, they read, $\tilde{b}= (+0.029, -0.007, -0.57 \ \textrm{Mpc})$ for $M, \Omega_k$ and $r_d$, respectively. In Sec. \ref{sec:res_fore}, we show that this bias is reduced as data quality improves, as expected, providing further evidence that it stems from the GP reconstruction itself.

\subsection{Forecasts}\label{sec:res_fore}

For the forecast analysis, we consider different combinations of current and future data sets -- e.g., current CCH with future SNIa -- to isolate the contribution of each probe to the overall constraints. This approach allows us to quantify how improvements in CCH, SNIa, or BAO individually affect the determination of $M$, $\Omega_k$, $r_d$, and $H_0$, including the role of redshift coverage in each sample. We also investigate the impact of data quality by considering pessimistic and optimistic scenarios for future CCH and SNIa samples, enabling the forecasts to highlight which improvements in the low- or high-redshift regimes are most influential.

We decide to present the forecast results in two separate tables, according to the pessimistic and optimistic scenarios for CCH and SNIa data (see Secs. \ref{sec:cch_fore} and \ref{sec:snIa_fore} for details). Our results are displayed in Tables \ref{tab:fore_pes} and \ref{tab:fore_opt}, respectively, where we report the upper and lower uncertainties at 68\% C.L. on $M$, $\Omega_k$, $r_d$, $H_0$ and $q_0$, together with the corresponding standard deviation, $\sigma$. A summary of the forecasted constraints obtained with the 2D analyses is provided in Figs. \ref{fig:2Dsncch} and \ref{fig:2Ddesicch}. The following comments are in order.

\begin{figure}
    \centering
    \includegraphics[width=\linewidth]{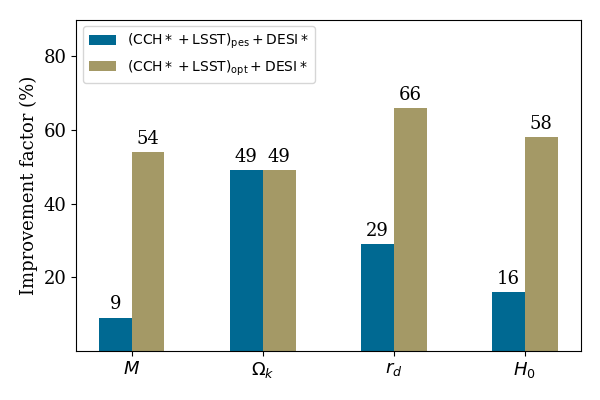}
    \caption{Percentage improvement factor on the parameters studied in this work ($x$-axis) achieved by future surveys in the pessimistic and optimistic scenarios for CCH$^{*}$+LSST-Y10+DESI$^{*}$ analyses. This is computed over the results obtained with the most recent available data, CCH+DESY5+DESI DR2.}\label{fig:imp_factor}
\end{figure}

\begin{itemize}
    \item \textit{Constraints on M}: the CCH+LSST-Y10 analysis shows no improvement in the estimation of $M$ compared to the CCH+DESY5 analysis, under both pessimistic and optimistic scenarios, since all of them lead to $\sigma_{M}\approx0.11$. However, this conclusion changes once current CCH data are replaced with the mock data set representing future measurements. In the pessimistic set-up, we obtain $\sigma_{M}=0.089$ with CCH$^{*}$+DESY5 and $\sigma_{M}=0.086$ with the full forecast CCH$^{*}$+LSSTY-10. When considering the optimistic scenario, these results change to $\sigma_{M}=0.057$ and $\sigma_{M}=0.049$, respectively. As already noticed in Sec. \ref{sec:res_curr}, within this approach, one can observe a substantial improvement in $M$ only with a decrease of the CCH error bars. This is why CCH$^{*}$ has such a big impact on the results. Remarkably, the full 3D joint analysis with CCH$^{*}$+LSST-Y10+DESI$^{*}$ leads to $\sigma_{M}=0.043$ in the optimistic scenario, improving by 54\% on the result obtained with CCH+DESY5+DESI DR2 (see Fig. \ref{fig:imp_factor}).

\begin{figure}
    \centering
    \includegraphics[width=\columnwidth]{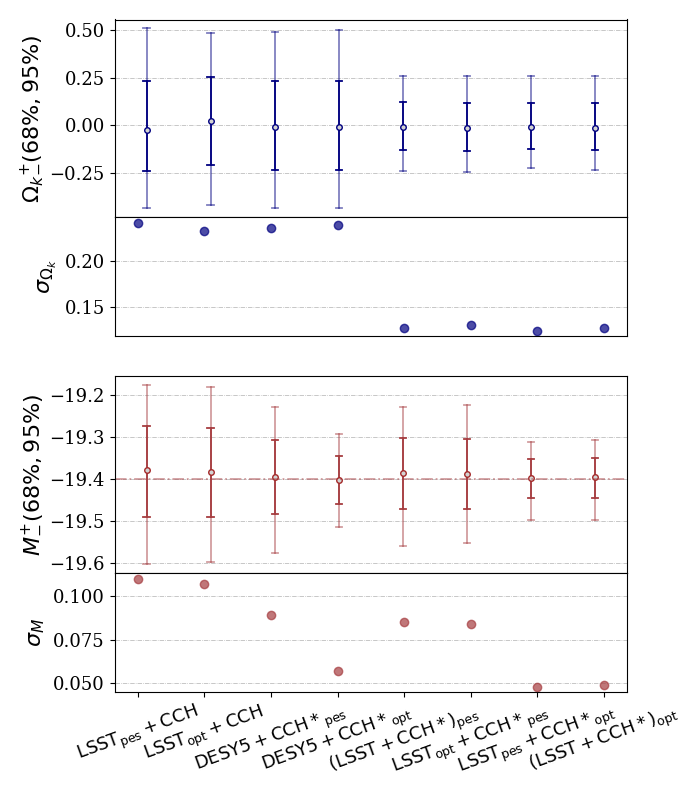}
    \caption{Error bars at 68\% and 95\% C.L. on the parameters $\Omega_k$ and $M$ obtained from the 2D joint analysis presented in Sec. \ref{sec:res_fore} with different combinations of CCH and SNIa data and under different scenarios (optimistic and/or pessimistic). Central values for each parameter are consistent with the $\Lambda$CDM prediction used to build the mock data, $\Omega_k=0$ and $M=-19.4$ (the latter indicated by the dot-dashed line in the bottom plot). Tiny fluctuations are due to the bias given by the GP reconstruction of the Hubble function (see Secs. \ref{sec:method_3} and \ref{sec:res_fore}). Lower panels of each plot show the standard deviation, $\sigma_{X}$, of the corresponding one-dimensional marginalized posterior.}
    \label{fig:2Dsncch}
\end{figure}%

\begin{figure}
    \centering
    \includegraphics[width=\columnwidth]{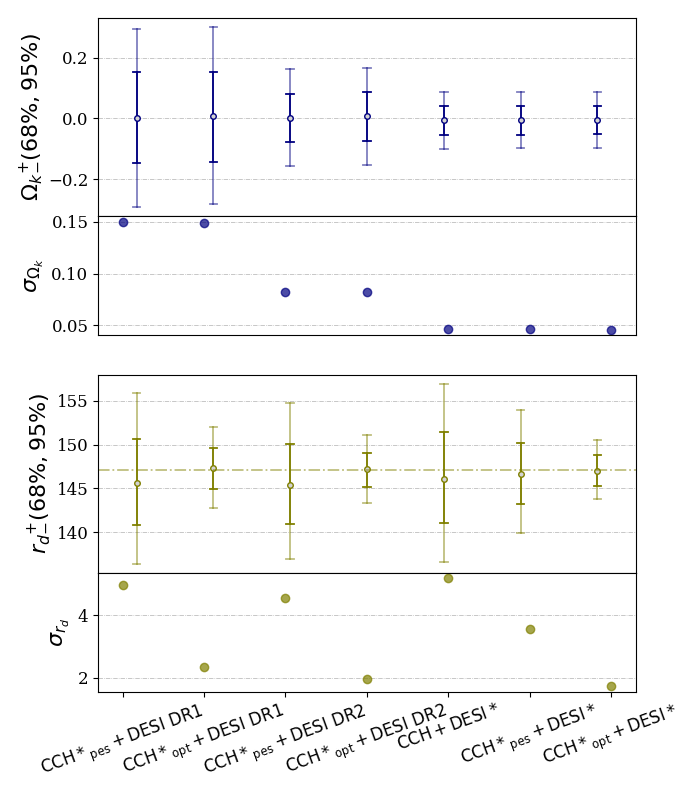}
    \caption{Same as in Fig. \ref{fig:2Dsncch} but for the parameters $\Omega_k$ and $r_d$ obtained from the 2D joint analysis presented in Sec. \ref{sec:res_fore} with different combinations of CCH and BAO data. In the bottom plot, the dot-dashed line represents the value of the sound horizon assumed to generate the mock data, $r_d=147.09$ Mpc. As in Fig. \ref{fig:2Dsncch}, fluctuations around this value reflect the intrinsic methodological bias introduced by the GP, which remains much smaller than the typical statistical uncertainty. See Sec. \ref{sec:res_fore} for details.}
    \label{fig:2Ddesicch}
\end{figure}%

    \item \textit{Constraints on $\Omega_k$:} the curvature parameter is the only parameter constrained by the two 2D analyses, since the luminosity and angular diameter distances entering the SNIa and BAO observables are sensitive to it. The increased statistics of LSST SNIa impact the constraint on $\Omega_k$ from the CCH+LSST-Y10 analysis, leading to a decrease of $\sim$20\% of the uncertainty compared to CCH+DESY5 ($\sigma_{\Omega_k}\approx0.30$), in both pessimistic and optimistic scenarios, for which $\sigma_{\Omega_k}=0.241$ and $\sigma_{\Omega_k}=0.231$, respectively. However, when CCH$^{*}$ are used, the improvement increases up to 58\% ($\sigma_{\Omega_k}=0.127$), meaning that a better treatment of the systematics affecting the CCH method will help tighten constraints on the curvature parameter. Furthermore, extended cosmic-clock galaxies observations up to $z\sim2.1$ will nearly double the data volume between $1.5 < z < 2.1$, thereby enhancing the constraining power in the regime where $\Omega_k$ is most sensitive.
    The role of CCH data is bypassed once we include in the analysis DESI data on anisotropic BAO (as already seen in Sec. \ref{sec:res_curr}), whose future measurements will be able to constrain $\Omega_k$ with $\sigma_{\Omega_k}=0.046$. The inclusion of SNIa slightly decreases this uncertainty to $\sigma_{\Omega_k}=0.044$, improving by a factor $\sim50$\% on current constraints (see Fig. \ref{fig:imp_factor}).
    The performance of future data to model-independently constrain the curvature parameter has been recently studied in, e.g., \cite{Matos:2023jkn, Hu:2024niv, Amendola:2024gkz, LHuillier:2024rmp}. For instance, our results are fully compatible with \cite{Hu:2024niv}, where authors obtained a precision of $\mathcal{O}(10^{-2})$ using time-delay distance from strong gravitational lenses, distance modulus measurements from LSST-like SNIa and $H(z)$ from future DESI BAO. With a different forecast analysis with DESI and Euclid data, \cite{Amendola:2024gkz} constrained the spatial curvature with $\sigma_{\Omega_k}=0.057$, by studying the deviations from statistical isotropy due to the Alcock-Paczyński effect of large-scale galaxy clustering. It will be difficult to constrain $\Omega_k$ better than few percent level with future data applying model-independent techniques.

    \item \textit{Constraints on $r_d$:} the sound horizon $r_d$ shows the most significant difference between pessimistic and optimistic scenarios concerning the CCH data. Its uncertainty ranges from $\sigma_{r_d} = 4.98$ (4.56) Mpc to $\sigma_{r_d} = 2.25$ (1.97) Mpc when considering CCH$^{*}$+DESI DR1 (DR2). This is essentially due to the fact that, on average, CCH uncertainties will decrease by a factor $\sim 70-80\%$ up to $z\sim2$, considering the systematics due to SPS models resolved (see Sec. \ref{sec:cch_fore}). $r_d$ is the parameter that shows the greatest improvement compared to constraints obtained with current data (see Fig. \ref{fig:imp_factor}). Within CCH$^{*}$+LSST-Y10+DESI$^{*}$ in the optimistic case, we are able to constrain $r_d$ with $\sigma_{r_d} = 1.73$ Mpc. This result is $\sim$80\% less precise than current model-dependent constraints. Indeed, in the context of standard pre-recombination physics, $r_d=(147.09\pm0.26)$ Mpc \citep{Aghanim:2018eyx}. Thus, our forecast corresponds to a 1.2\% precision on the fiducial sound horizon. This result improves upon previous model-independent forecasts, which, however, have been obtained using different combinations of data sets and methodology. For instance, \cite{Giare:2024syw} show that a 1.5\% precision measurement of $r_d$ will be possible with future standard sirens observed with LISA \citep{LISACosmologyWorkingGroup:2019mwx} in combination with angular BAO measurements from DESI or Euclid.
    
    \item \textit{Constraints on $H_0$:} once obtained our model-independent constraint on $M$, we can break the degeneracy between the latter and $H_0$ to obtain an estimate of the Hubble parameter using the local distance ladder. By virtue of Eq. \eqref{eq:H0}, and using the LSST SNIa in the Hubble flow, we obtain $\sigma_{H_0} = 2.7$ km/s/Mpc and $\sigma_{H_0} = 1.35$ km/s/Mpc for the pessimistic and optimistic scenario, respectively, and the following uncertainities on the deceleration parameter, $\sigma_{q_0} = 0.06$ and $\sigma_{q_0} = 0.04$. Considering having SNIa and CCH systematics well-understood and under control in the next decade, this means constraining $H_0$ at 2\% level without relying on any cosmological model. A very similar result is obtained in forecast analyses performed in \cite{Matos:2023jkn}, although with a different methodology and combination of probes, which includes, for instance, the use of standard sirens. Future CCH alone will obtain the same level of precision, but in the context of a concrete cosmological model \citep{Moresco:2024wmr}. 
    As for the deceleration parameter, the improvement factor over the typical error obtained with DESY5 in the cosmographical fit, $\sigma_{q_0} \approx 0.32$ (cf. Table \ref{tab:res_curr}), is $\sim 8$, essentially driven by the substantial increase in the number of SNIa in the Hubble flow.
    
\end{itemize}
For completeness, we also performed all the possible combinations of scenarios between the data sets considered, e.g. LSST-Y10$_{\rm opt}$+CCH$^{*}_{\rm pes}$. Since the conclusions that can be drawn from these intermediate analyses only reinforce those discussed above, we just refer the reader to the results displayed in Figs. \ref{fig:2Dsncch} and \ref{fig:2Ddesicch}. Essentially, the replacement of LSST-Y10$_{\rm pes}$ with LSST-Y10$_{\rm opt}$ only leads to a very marginal decrease of the error bars, which is almost unnoticeable, while the replacement of CCH$^{*}_{\rm pes}$ with CCH$^{*}_{\rm opt}$ induce a significant improvement in the constraining power.
From these figures, we also observe that the GP reconstruction bias decreases as data precision increases. For a given CCH sample, the offset remains nearly constant across scenarios, further indicating that this systematic shift is an intrinsic methodological feature inherent to the GP approach.
We conclude that improving the data quality reduces the magnitude of $|\tilde{b}|$, which tends to vanish in the limit of nearly noise-free data, consistent with the definition of the GP posterior mean (see Sec. \ref{sec:method_2}). In the optimistic scenario, the offsets observed are typically $\lesssim0.1\sigma$ in absolute value in all cases. It is worth emphasizing that our framework enables the simultaneous estimation of both statistical uncertainties and the typical level of methodological bias associated with the adopted reconstruction technique.

As a final remark, we expect that all the results presented in this section would benefit from the combination of current and future measurements (see, e.g. \cite{Moresco:2024wmr} concerning CCH data), something that we avoid exploring in this work as it would require a treatment of correlations between samples.
Additionally, beyond LSST, the Roman Space Telescope \citep{2015arXiv150303757S} may offer a valuable opportunity to reassess the results of this work, since it will be able to detect SNIa up to $z\sim3$, which may significantly improve the estimation of $\Omega_k$, in particular.

\begin{table*}
\centering
\caption{Errors at 68\% C.L. on the ladder calibrators, $M$ and $r_d$, and on the curvature parameter $\Omega_k$ entering Eq. \eqref{eq:DLx}, from the joint analyses presented in Sec. \ref{sec:res_fore} considering the \textit{pessimistic} scenario for future data on CCH and SNIa. In the last two columns, we report the cosmographical constraints on $H_0$ and $q_0$, cf. Eq. \eqref{eq:H0}. For each parameter, in the parenthesis we also report the standard deviation ($\sigma$) of the corresponding one-dimensional probability posterior.}\label{tab:fore_pes}
\begin{tabular}{lrrrrr}
&$M$ & $\Omega_k$ & $r_d$ [Mpc] & $H_0$ [km/s/Mpc] & $q_0$\\ \hline\hline
CCH+\textbf{LSST-Y10} & $_{-0.112}^{+0.106}$ (0.110) & $_{-0.219}^{+0.254}$ (0.241) &-& - & -\\
\hline
\textbf{CCH$^{*}$}+DESY5 & $_{-0.090}^{+0.087}$ (0.089)& $_{-0.222}^{+0.242}$ (0.235)&-& - & -\\
\textbf{CCH$^{*}$}+DESI DR1 &-& $\pm0.150$ (0.150) &$_{-4.78}^{+5.08}$ (4.98) & - & -\\
\textbf{CCH$^{*}$}+DESI DR2 &-&$\pm0.080$ (0.082)& $_{-4.38}^{+4.68}$ (4.56)& - & -\\
\hline
CCH+\textbf{DESI$^{*}$} &- &$\pm0.047$ (0.047)&$_{-4.95}^{+5.37}$ (5.18)& - & -\\
\hline\hline
\textbf{CCH$^{*}$+LSST-Y10} &$_{-0.088}^{+0.082}$ (0.086) & $_{-0.121}^{+0.130}$ (0.127) &-&- & -\\
\hline
\textbf{CCH$^{*}$+DESI$^{*}$}&-&$\pm0.047$ (0.047)& $_{-3.49}^{+3.56}$ (3.57) & - & -\\
\hline
\textbf{CCH$^{*}$+LSST-Y10+DESI$^{*}$} &$_{-0.082}^{+0.087} (0.085)$ & $_{-0.043}^{+0.045}$ (0.044)& $_{-3.49}^{+3.56}$ (3.57) & $_{-2.82}^{+2.54}$ (2.70) & 0.06\\
\hline
\end{tabular}%
\end{table*}

\begin{table*}
\centering
\caption{Same as in Table \ref{tab:fore_pes} but considering the \textit{optimistic} scenario. Notice that the row corresponding to the CCH+DESI$^{*}$ analysis is the same as the previous table, as no distinction between scenarios is adopted for DESI data.}\label{tab:fore_opt}
\begin{tabular}{lrrrrr}
&$M$ & $\Omega_k$ & $r_d$ [Mpc] & $H_0$ [km/s/Mpc] & $q_0$\\ \hline\hline
CCH+\textbf{LSST-Y10} & $_{-0.111}^{+0.095}$ (0.105) & $_{-0.203}^{+0.246}$ (0.231)&-&-\\
\hline
\textbf{CCH$^{*}$}+DESY5 & $0.056$ (0.057) &$_{-0.226}^{+0.246}$ (0.239) &-& - & -\\
\textbf{CCH$^{*}$}+DESI DR1 &-& $_{-0.150}^{+0.145}$ (0.149) & $_{-2.39}^{+2.29}$ (2.25) & - & -\\
\textbf{CCH$^{*}$}+DESI DR2 &-&$\pm0.08$ (0.082)&$_{-1.99}^{+1.89}$ (1.97)& - & -\\
\hline
CCH+\textbf{DESI$^{*}$} &- &$\pm0.047$ (0.047)&$_{-4.95}^{+5.37}$ (5.18)& - & -\\
\hline\hline
\textbf{CCH$^{*}$+LSST-Y10} &$_{-0.050}^{+0.046}$ (0.049)& $_{-0.118}^{+0.132}$ (0.127)&-&- & -\\
\hline
\textbf{CCH$^{*}$+DESI$^{*}$} &-&$_{-0.045}^{+0.047}$ (0.046)& $_{-1.70}^{+1.75}$ (1.73)& - & -\\
\hline
\textbf{CCH$^{*}$+LSST-Y10+DESI$^{*}$} &$_{-0.040}^{+0.045}$ (0.043) & $_{-0.045}^{+0.043}$ (0.044) & $_{-1.70}^{+1.75}$ (1.73) & $_{-1.34}^{+1.37}$ (1.35) & 0.04\\
\hline
\end{tabular}%
\end{table*}

\section{Conclusions}\label{sec:conclusions}
The role that the calibrators of the local and inverse distance ladder, $M$ and $r_d$, have in measuring cosmic distances across our universe is crucial. Potential unknown systematics or hidden model dependencies may bias our understanding of fundamental cosmological properties and impact the discourse on the Hubble tension in a dangerous way. In this context, it is of utmost importance to measure these quantities following a methodology as model-independent as possible. On the other hand, having also an accurate and agnostic determination of the spatial curvature becomes fundamental, as it has important implications for our understanding of the inflationary epoch. To this end, the contribution of low-redshift data gathered by current surveys and telescopes is becoming increasingly valuable, and we anticipate even a larger impact from future large-volume and high-precision observations. Building on the model-independent calibration of the cosmic ladders introduced in our previous work, \cite{Favale:2023lnp}, in this manuscript, we have performed a follow-up analysis by incorporating the most recent data on supernovae of type Ia and baryon acoustic oscillations as provided by the DES and DESI collaborations. Our approach continues to rely on state-of-the-art data on cosmic chronometers up to redshift $\sim$2. With the aid of the Gaussian Processes technique, we have applied a quite model-independent methodology to simultaneously constrain $M$, $\Omega_k$ and $r_d$. We have done so by using data free from the main drivers of the current $\gtrsim5\sigma$ tension on $H_0$ -- the first rungs of the local distance ladder employed by SH0ES and CMB data from \textit{Planck}. Under the umbrella of CCH, SNIa from DESY5 and BAO from DESI DR1, we obtain $M =-19.330_{-0.099}^{+0.095}$, $\Omega_k = -0.113_{-0.155}^{+0.150}$ and $r_d = (145.30_{-5.08}^{+5.28})$ Mpc. Using the most recent and more precise DESI DR2 data, we have been able to place tighter constraints on these quantities, $M = -19.324_{-0.095}^{+0.092}$, $\Omega_k = -0.143\pm0.085$ and $r_d = (144.00_{-4.88}^{+5.08})$ Mpc, which improve those provided in our previous work. The CCH+DESY5+DESI DR2 constraint on $M$ and the DESY5 SNIa in the Hubble flow prefer a Hubble parameter closer to the CMB-inferred value, being $H_0 = (68.83_{-3.07}^{+3.03})$ km/s/Mpc. Remarkably, substituting $M$ with the SH0ES value ($M^{R22}$), we obtain $H_0 = (70.84\pm1.16)$ km/s/Mpc, reducing the tension with CMB estimates below 3$\sigma$. This may have implications for the discussion about possible systematics in low-$z$ SNIa that could bias the local estimate of $H_0$.
We have also studied the case in which DESY5 SNIa are replaced with Pantheon+, i.e. CCH+PAN+DESI DR2. This slightly decreases the uncertainties on $M$, $\Omega_k$ and $r_d$ by 6\%, 5\% and 1\%, respectively, while the corresponding cosmographic estimate of $H_0$ reads $H_0 = (72.12_{-2.95}^{+2.97})$ km/s/Mpc. 
However, current observations still limit the precision on these quantities and the conclusions that can be drawn from them. Another example is the result obtained for $\Omega_k$, which points towards a positive curvature ($\Omega_k<0$). However, with the full combination of data, our constraint is still compatible with the flatness assumption at 1.7$\sigma$ when using DES SNIa and at 1.2$\sigma$ when the latter is replaced with Pantheon+.

The robustness of these constraints has been assessed by running the analysis pipeline on mock data that incorporate current observational uncertainties. This allowed us to validate the entire pipeline and test the methodological assumptions. In particular, we quantified the impact of a systematic methodological bias inherent to the GP technique on the inferred parameters, finding the largest offset on $M$ ($\sim0.3\sigma$), and smaller biases on the remaining parameters ($\lesssim0.1\sigma$).

We have then extended our analysis with forecasts on future data, which will be provided by LSST, Euclid, WST, ATLAS and the final-year DESI observations.
Our main results highlight that, in a pessimistic scenario, the combination CCH$^{*}$+LSST-Y10+DESI$^{*}$ will constrain the SNIa absolute magnitude and the sound horizon with $\sigma_{M} = 0.085$ and $\sigma_{r_d} = 3.57$ Mpc, respectively. This is mainly driven by the conservative treatment of systematics in the cosmic chronometers method, since CCH act as our calibrator of the ladders. On the curvature parameter, we obtain an error of $\sigma_{\Omega_k} = 0.044$.
The optimistic scenario leads instead to $\sigma_{M} = 0.043$, $\sigma_{r_d} = 1.73$ Mpc and $\sigma_{\Omega_k} = 0.044$. This translates into an improvement by a factor of 54\%, 66\% and 49\%, respectively, on current constraints obtained with the most recent data (CCH+DESY5+DESI DR2). Thus, up to $z\approx2$, we forecast that the relative precision on the SNIa absolute magnitude will be $\sigma_{M}/M \sim 0.2\% $, which allows for constraining the Hubble parameter at a 2\% level, being $\sigma_{H_0} = 1.35$ km/s/Mpc. The relative precision on the sound horizon, $\sigma_{r_d}/r_d \sim 1.2\%$, will instead be valuable for assessing early-time solutions to the Hubble tension, as those which require new physics prior to recombination that would reduce the value of $r_d$. The presented results are robust against the intrinsic methodological bias, which remains below $0.1\sigma$ for all parameters thanks to the increased precision of future data.
In conclusion, this first forecast analysis on the presented methodology, though not yet as precise as model-dependent analyses, is already nearly competitive with current constraints derived from those methods, thereby paving the way for model-independent constraints expected in the coming decade. It provides valuable insight into where improvements are most needed -- such as redshift coverage, data volume and assessment of systematics. As already highlighted, future missions beyond those used in this work could prove essential for a thorough re-evaluation of our findings. This could be the case of the Roman Space Telescope, which is expected to considerably extend the detection of SNIa up to $z\sim3$. This broader redshift coverage will allow for a more robust anchoring of the Hubble diagram and improved sensitivity to cosmic curvature, enabling tighter tests of the distance ladder and potential deviations from $\Lambda$CDM. In this sense, the methodology explored in this paper offers a timely and flexible tool to interpret and cross-check future high-precision cosmological data.

\section*{Acknowledgements}
We thank Oscar Straniero for the useful discussions that motivated us to carry out the forecast analysis presented in this work. The authors thank the anonymous referee for their insightful comments and suggestions. AF and MM acknowledge support from the INFN project “InDark”.
AGV is funded by “la Caixa” Foundation (ID 100010434) and the European Union’s Horizon 2020 research and innovation programme under the Marie Sklodowska-Curie grant agreement No 847648, with fellowship code LCF/BQ/PI23/11970027. We are supported by the
ASI/LiteBIRD grant n. 2020-9-HH.0 and by the Italian Research Center on High Performance Computing Big Data and Quantum Computing (ICSC), project funded by European Union - NextGenerationEU - and National Recovery and Resilience Plan (NRRP) - Mission 4 Component 2 within the activities of Spoke 3 (Astrophysics and Cosmos Observations). AF and AGV acknowledge the participation in the COST Action CA21136 “Addressing observational tensions in cosmology with systematics and fundamental physics” (CosmoVerse).
AF is grateful to the Institute of Cosmos Sciences of the University of Barcelona for their hospitality during the writing of this manuscript.

\section*{Data Availability}

The data employed in this article are publicly available (see Sec. \ref{sec:data} and references therein).


\bibliographystyle{mnras}
\bibliography{amain}


\appendix

\section{GP kernel sensitivity test}\label{app:kernel}
In this Appendix, we briefly report the results of the GP kernel sensitivity test for the analysis of Sec. \ref{sec:res_curr} with current data, including DESY5 and DESI DR2. The goal is to assess the robustness of our conclusions under a different choice of GP kernel, as in \cite{Favale:2023lnp}. While our baseline analysis employs Matérn 3/2 (M32), which is a once differentiable kernel, we opt to repeat the GP reconstruction using the Matérn 5/2 (M52) kernel, which is twice differentiable. Limited kernel differentiability can indeed affect derivative constraints near the boundaries of the reconstruction, a feature particularly relevant for $\Omega_k$, which enters through a nonlinear function of the comoving distance and can be more sensitive to small, coherent changes in the reconstructed expansion history.
From the 3D joint analysis based on the M52 reconstructions of $H(z)$, we obtain $M=-19.303_{-0.093}^{+0.090}$, $\Omega_k=-0.128\pm0.080$ and $r_d=144.20_{-4.78}^{+5.08}$ Mpc. Comparing these results with those of Sec. \ref{sec:res_curr} obtained with M32, we find that all parameters are fully consistent within 0.2$\sigma$, confirming the robustness of our conclusions. Only the CCH+DESY5 result for $\Omega_k$ shows a slightly larger deviation (0.7$\sigma$), being $\Omega_k=-0.257_{-0.264}^{+0.304}$, which leads to a $\sim0.9\sigma$ departure from the flatness assumption. However, when the full data combination is considered, we still find a 1.6$\sigma$ departure, the same as the baseline M32 result. This stability arises because BAO data, which measure distances in units of $r_d$, absorb part of the kernel-induced variation in the reconstructed expansion history, preventing these shifts from propagating into $\Omega_k$ and reducing its sensitivity to the GP kernel choice. See Table \ref{tab:M52} for the complete list of the results obtained with M52.

\begin{table}
\centering
\caption{Constraints at 68\% C.L. on $M$, $\Omega_k$ and $r_d$ obtained using the GP reconstructions from the M52 kernel. For a direct comparison, we additionally report the results obtained in Sec. \ref{sec:res_curr} with M32.}\label{tab:M52}
\resizebox{\columnwidth}{!}{%
\begin{tabular}{lrrr}
& CCH+DESY5 & CCH+DESI DR2 & CCH+DESY5+DESI DR2 \\ \hline\hline
\multicolumn{4}{c}{M52}\\ \hline
M &   $-19.286_{-0.108}^{+0.104}$ & - & $-19.303_{-0.093}^{+0.090}$ \\
$\Omega_k$ & $-0.257_{-0.264}^{+0.304}$ &  $-0.118_{-0.080}^{+0.085}$ &  $-0.128\pm0.080$\\
$r_d$ [Mpc] & - & $144.30_{-4.78}^{+5.18}$ & $144.20_{-4.78}^{+5.08}$ \\
\hline
\multicolumn{4}{c}{M32}\\ \hline
M & $-19.277_{-0.106}^{+0.104}$ & - & $-19.324_{-0.095}^{+0.092}$ \\
$\Omega_k$ & $-0.48_{-0.27}^{+0.32}$ & $-0.12\pm0.09$ &  $-0.143\pm0.085$\\
$r_d$ [Mpc] & - & $144.20_{-4.88}^{+5.18}$ & $144.00_{-4.88}^{+5.08}$ \\
\hline
\end{tabular}
}
\end{table}

\section{Fit of the posteriors}\label{app:fit}
\begin{figure}
    \centering
    \includegraphics[width=0.8\linewidth]{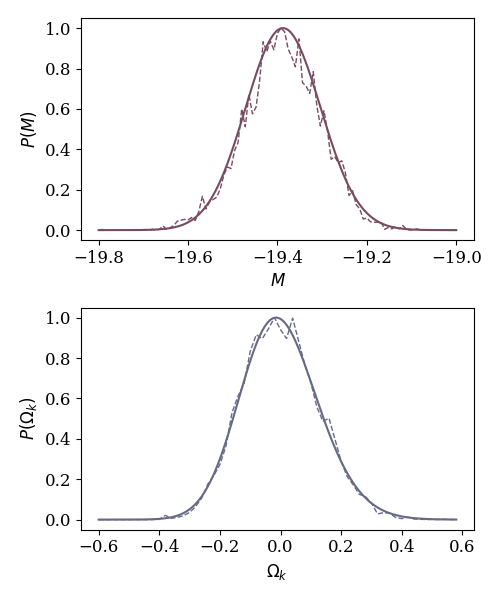}
    \caption{Example of the application of the skewed-normal fit (solid lines) on the noisy posteriors (dashed lines) for $M$ and $\Omega_k$ from the 2D joint analysis with LSST$_{\rm opt}$ + CCH$^{*}_{\rm pes}$. In this case, we obtain $\alpha_M = -1.09$ and $\alpha_{\Omega_k} = 1.16$.}
    \label{fig:fitpost}
\end{figure}

As mentioned in Sec. \ref{sec:method}, in the context of the forecast analysis, it may happen to have only a limited number of $N$ realizations of $H(z)$ that can be computed to obtain Eq. \ref{eq:wmu}, as a consequence of long runtime requirements. This yields the corresponding distributions for $M$, $\Omega_k$ and $r_d$ to have some degree of noise. We smooth it out by fitting a skew-normal probability density function (PDF) to the one-dimensional posterior probability for each parameter. The choice of this particular function lies in the typical shape of the posteriors observed in both the previous analysis \citep{Favale:2023lnp} and the updated results in Sec. \ref{sec:res_curr} (see Fig. \ref{fig:triangleDR2}).
The skewed-normal function accounts for small deviations from the Gaussian distribution, i.e. for skewness $\neq 0$.
The PDF of this distribution takes the form:
\begin{equation}\label{eq:skew}
    f(x;\alpha,\nu,\delta) = \frac{2}{\delta}\cdot \phi\Big(\frac{x - \nu}{\delta}\Big)\cdot\Phi\Big(\alpha\cdot \frac{x - \nu}{\delta}\Big)\,,
\end{equation}%
where $\phi(x)$ is the standard normal PDF and $\Phi(x)$ its cumulative density function. The distribution is characterised by the triad $(\alpha, \nu, \delta)$, which are related, respectively, to the shape (skewness), location and scale parameters. When $\alpha=0$, one recovers the normal distribution. We obtain $-1.7<\alpha<1.9$ for all the cases under study, positive or negative depending on whether we have a left- or right-skewed density function. As expected, these values of $\alpha$ point to only a mild deviation from the Gaussian distribution, with the effect being more pronounced in the posteriors of $M$ and $\Omega_k$, which also show some degree of mutual correlation (see again Fig. \ref{fig:triangleDR2}). Indeed, considering the CCH+BAO analysis, we always find $\alpha < 1.2$, whereas for the 3D joint analyses $\alpha < 1 $. We report one example of this fit in Fig. \ref{fig:fitpost}.

\section{Pantheon+ vs. DESY5 Hubble flow}\label{app:HF}
\begin{figure}
    \centering
    \includegraphics[width=0.9\linewidth]{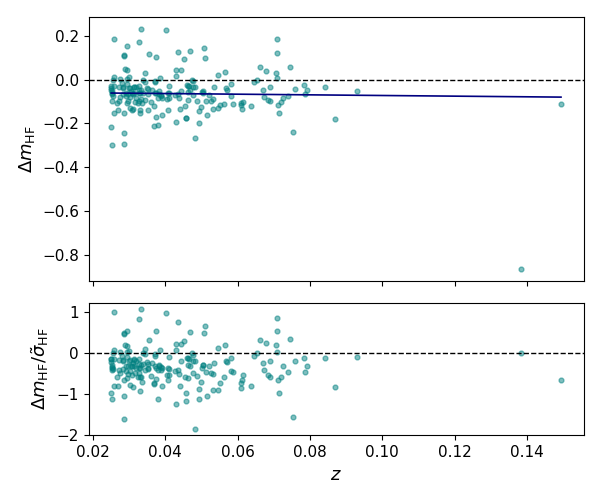}
    \caption{Residuals of the apparent magnitude $m$ as a function of the redshift, $\Delta m_{\rm HF}$, for the SNIa common to Pantheon+ and DESY5 samples in the Hubble flow (HF). The large $\Delta m_{\rm HF}$ observed at $z\sim0.14$ is associated with a SNIa in the DESY5 sample with uncertainty $\delta m>1$, and therefore contributes negligibly, as reflected in the weighted residuals, which are shown in the bottom panel. The solid line in the upper panel represents a linear fit to the residuals and is shown as a visual guide to assess the presence of a constant offset or a redshift-dependent trend.}
    \label{fig:HF}
\end{figure}

In Sec. \ref{sec:res_curr}, we showed that a lower value of $H_0$ is preferred by the DESY5 Hubble flow, even when adopting the SH0ES prior on $M$, and that this shift may originate from systematic differences between the Pantheon+ and DESY5 Hubble flow regimes. To complement that discussion, in this Appendix, we compare these SNIa by selecting objects common to both compilations in the region $0.023\lesssim z \lesssim 0.15$. In particular, in Fig. \ref{fig:HF}, we show the residuals of the apparent magnitudes, $\Delta m_{\rm HF} = (m_{\rm PAN} - m_{\rm DES})_{\rm HF}$, together with the corresponding residuals weighted by the standard deviation, $\tilde{\sigma}_{\rm HF} = \sqrt{\sigma_{\rm PAN}^{2} + \sigma_{\rm DES}^{2}}$. This allows us to observe that, while the majority of the residuals lie within 1$\sigma$, there is a clear trend toward $\Delta m_{\rm HF}<0$, indicating that DESY5 SNIa are systematically fainter than the same events in the Pantheon+ sample, thereby biasing distance estimates and shifting the inferred Hubble parameter. These differences can be driven by methodological choices -- such as bias corrections, intrinsic scatter models, or SNIa selection criteria -- although a detailed assessment of their origin lies beyond the scope of this work. For recent discussions on these matters, we refer the reader to, e.g., \cite{Efstathiou:2024xcq, DES:2025tir}.

%


\bsp	
\label{lastpage}
\end{document}